\documentclass{elsart}
\usepackage{amssymb}
\usepackage{epsfig}
 \usepackage{amssymb}
 \usepackage{epsf}
 \usepackage{epsfig}
 \usepackage{color}
    \newcommand{\ba}{\begin{eqnarray}}
    \newcommand{\ea}{\end{eqnarray}}
    \newcommand{\be}{\begin{equation}}
    \newcommand{\ee}{\end{equation}}

    \newcommand {\bp} {{\mathbf p}}

    \newcommand{\AmS}{{\protect\the\textfont2%
  A\kern-.1667em\lower.5ex\hbox{M}\kern-.125emS}}

\newcommand{\calD}{{\mathcal D}}

\newcommand{\calN}{{\mathcal N}}
\newcommand{\calU}{{\mathcal U}}
\newcommand{\calV}{{\mathcal V}}

\newcommand {\fslash}[1]{{#1\kern -0.45em /\kern 0.3em}}

\begin{document}
\runauthor{PKU}
\begin{frontmatter}

\title{Massive Domain Wall Fermions on Four-dimensional Anisotropic Lattices\thanksref{fund}}


\author[PKU]{Xu Feng},
\author[PKU]{Xin Li},
\author[PKU]{Wei Liu}
\author[PKU]{Chuan Liu},
\address[PKU]{School of Physics, Peking University\\
              Beijing, 100871, P.~R.~China}

\thanks[fund]{This work is supported by the Key Project of National Natural
Science Foundation of China (NSFC) under grant No. 10235040, No.
10421003, and supported by the Trans-century fund and the Key
Grant Project of Chinese Ministry of Education (No. 305001).}

\begin{abstract}
 We formulate the massive domain
 wall fermions on anisotropic lattices.
 For the massive domain wall fermion, we find that the
 dispersion relation assumes the usual form in
 the low momentum region when the bare parameters
 are properly tuned.
 The quark self-energy
 and the quark field renormalization constants are
 calculated to one-loop in bare lattice perturbation theory.
 For light domain wall
 fermions, we verified that the chiral mode is stable against
 quantum fluctuations on anisotropic lattices.
  This calculation serves as a guidance for the tuning of
 the parameters in the quark action in future numerical simulations.
\end{abstract}
\end{frontmatter}


\section{Introduction}

 In recent years, considerable progress has been made in
 understanding chiral symmetry on the lattice. Domain wall
 fermions~\cite{kaplan92:DWF_idea,shamir93:DWF_boundary} and
 the overlap fermions~\cite{neuberger93:overlap_prl,neuberger94:overlap_npb,neuberger98:massless_vector,neuberger98:massless_quark}
 have emerged as two new
 candidates in the formulation of lattice fermions which
 have much better chiral properties than the conventional
 lattice Wilson or staggered fermions. Since chiral symmetry
 is so crucial to the theory of QCD, it is therefore desirable
 to use these new fermions if possible.
 However, due to their more expensive computational cost, this
 task has not been fully accomplished, particularly for the
 study of multi-hadron states and hadrons with massive quarks.

 On the other hand, anisotropic lattices have been used to
 study heavy hadronic states and they proved to be extremely
 helpful in various applications. These include: glueball
 spectrum calculations~\cite{colin99,chuan01:gluea},
 charmonium spectrum calculations~\cite{chen01:aniso,CPPACS02:aniso},
 charmed meson and charmed baryon calculations~\cite{juettner03:Ds,lewis01:aniso}
 and hadron-hadron scattering
 calculations~\cite{chuan02:pipiI2,chuan04:KN,chuan04:Kpi,chuan04:pipi}.
 Many of these studies in fact involve light quarks and chiral symmetry
 plays an essential role. It is therefore more appropriate to
 study the light quarks using lattice fermions with
 better chiral properties.
 Using domain wall fermions with
 the physical four-dimensional lattice being isotropic, the
 heavy-light systems~\cite{RBC05:DWF_heavy_light} and
 hadron-hadron scattering~\cite{savage06:pipi_dwf} have been studied.
 As we pointed out, anisotropic lattices can be very helpful in these studies.
 It is therefore desirable
 to study domain wall fermions with the four-dimensional
 physical lattice being anisotropic and this will
 be the major purpose of this paper.
 Major possible applications of the anisotropic domain wall fermion
 approach that we have in mind are the heavy-light hadrons, exotic
 hadrons with light quarks
 and hadron-hadron scattering processes where the
 objects being studied are heavy and chiral property is crucial.

 In this paper, we study an explicit formulation
 of domain wall fermions on anisotropic four-dimensional lattices.
 We adopt the domain wall fermion action which
 is of Shamir type~\cite{shamir93:DWF_boundary}.
 For the gauge action, we use
 the tadpole improved gauge actions~\cite{colin97,colin99} which have
 been used in glueball calculations.
 Similar to the case of Wilson fermions on anisotropic lattices,
 domain wall fermion action on anisotropic lattices contains
 more parameters than its isotropic counterparts. Some of these
 parameters have to be tuned properly in order to exhibit a correct
 continuum limit. This problem is first analyzed in the case
 of free domain wall fermions on anisotropic lattices. We find that,
 in order to restore the normal relativistic dispersion relation for
 the quark, parameters of the fermion action has to be tuned accordingly. Then,
 we compute the quark propagator in lattice perturbation theory
 to one-loop. Quark field and quark mass  renormalization
 constants are obtained for various values of the bare parameters.
 We also discuss the choice of the parameters which maintains
 good chiral symmetry. This perturbative calculation
 serves as a guidance for further non-perturbative Monte Carlo simulations
 which are currently under investigation~\cite{chuan06:future}.
 Similar perturbative calculations have been performed
 in the case of isotropic lattice~\cite{neuberger98:residual_mass,aoki99:DWF_pert_selfenergy}.
 The calculation in this paper is an extension to the anisotropic lattice.

 This paper is organized as follows. In section 2,
 domain wall fermion action on anisotropic lattices is given.
 In section 3, the free domain wall fermion propagator on anisotropic lattices
 is derived. We also study the dispersion relation of the
 free domain wall fermion on anisotropic lattices.
 It turns out that even in the free case, hopping parameters
 of the fermion action have to be tuned properly, according
 to the value of the quark mass, so as to have the correct
 continuum limit for the quark.
 In section 4, the strategies for the
 calculations of fermion self-energy to one-loop
 are outlined. In section 5, numerical results for
 the one-loop calculation are presented for various bare parameters.
 The renormalization factors for the quark field and
 the current quark mass are given.
 We also discuss the renormalization of parameter $M_5$
 within tadpole and mean-field approximation.
 This parameter has to be tuned to the right range in order
 to maintain chiral properties of the fermion.
 In section 6, we will conclude with some remarks and outlook.
 Dispersion relation of the free domain wall fermion
 is discussed in Appendix A.
 Some explicit formulae for the one-loop calculations
 are listed in Appendix B.

 \section{Lattice Action for Domain Wall Fermions on Anisotropic Lattices}
 \label{sec:lattice_action}

 The lattice gauge action  used in this study is the tadpole
 improved action on anisotropic lattices:~\cite{colin97,colin99}
 \ba
 \label{eq:gauge_action}
 S=&-&\beta\sum_{i>j} \left[
 {5\over 9}{TrP_{ij} \over \chi u^4_s}
 -{1\over 36}{TrR_{ij} \over \chi u^6_s}
 -{1\over 36}{TrR_{ji} \over \chi u^6_s} \right] \nonumber \\
 &-&\beta\sum_{i} \left[ {4\over 9}{\chi TrP_{0i} \over  u^2_s}
 -{1\over 36}{\chi TrR_{i0} \over u^4_s} \right] \;\;,
 \ea
 where $P_{0i}$ and $P_{ij}$ represents the usual temporal and
 spatial plaquette variable, respectively.
 $R_{ij}$ and $R_{i0}$ designates the $2\times 1$
 spatial and temporal Wilson loops, where, in order to eliminate
 the spurious states, we have restricted the coupling of fields
 in the temporal direction
 to be within one lattice spacing. The parameter $u_s$,
 which is usually taken to be the fourth root of the average
 spatial plaquette value in the simulation, implements the tadpole improvement.
 Using this gauge action, glueball and hadron spectra have been studied
 within quenched approximation \cite{colin97,colin99,%
 chuan01:gluea,chuan01:glueb,chuan01:canton1,chuan01:canton2,chuan01:india}.

 For the fermion action, we adopt Shamir type domain
 wall fermions~\cite{shamir93:DWF_boundary}.
 The five-dimensional fermion and anti-fermion field is denoted as $\psi_{x,s}$
 and $\bar{\psi}_{x,s}$, respectively.
 The fermion lattice action we studied is given by:
 \ba
 \label{eq:fermion_action}
 S_f &=& S'_f+S_m \;,
 \nonumber \\
 S'_f&=&-\sum_{x,s}(1+3\kappa_s+\kappa_t-M_5)\bar{\psi}_{x,s}\psi_{x,s}
 \nonumber \\
 &+&\sum_{x,s}\left[
 \bar{\psi}_{x,s}P_R\psi_{x,s+1}+\bar{\psi}_{x,s}P_L\psi_{x,s-1}
 \right]
 \nonumber \\
 &+&{1\over 2}\sum_{x,s,\mu}\kappa_\mu\bar{\psi}_{x,s}
 \left[(1+\gamma_\mu)U_\mu(x)\psi_{x+\hat{\mu},s}
 +(1-\gamma_\mu)U^\dagger_\mu(x-\hat{\mu})\psi_{x-\hat{\mu},s}\right]
 \;,\nonumber \\
 S_m &=& -m \sum_{x}\left[
  \bar{\psi}_{x,0}P_L\psi_{x,L_s-1}
 +\bar{\psi}_{x,L_s-1}P_R\psi_{x,0}
 \right]\;,
 \ea
 We will also use the notation $\kappa_0=\kappa_t$, $\kappa_i=\kappa_s$.
 The left and right-handed projectors are: $P_{L/R}=(1\mp\gamma_5)/2$.
 For definiteness, we take the extension in the fifth dimension
 to be $L_s$ and the coordinate of the fifth dimension is
 labelled such that $0\le s \le L_s-1$.
 Thus, our domain wall fermion action is
 characterized by four parameters: five-dimensional mass
 (wall height) parameter $M_5$,
 temporal hopping parameter $\kappa_t$, spatial
 hopping parameter $\kappa_s$ and current quark
 mass parameter $m$.
 \footnote{Note that there is also an additional hidden parameter
 in the theory, namely the extent of the fifth dimension: $L_s$}.

 \section{Free Domain Wall Fermion Propagator on Anisotropic Lattices}
 \label{sec:free_G}

 In this section, we will briefly review the results for
 the free propagator of the domain wall fermions on anisotropic lattices.
 Free domain wall fermion propagator on an isotropic lattice has been discussed in the
 literature~\cite{shamir93:DWF_boundary,shamir93:DWF_spectrum,neuberger98:residual_mass,aoki99:DWF_pert_selfenergy}.
 Our notations follow those in~\cite{shamir93:DWF_boundary,aoki99:DWF_pert_selfenergy}
 and we have made modifications to the anisotropic lattice where necessary.
 We have also tried to keep all relevant mass terms (bare quark mass $m$
 and the residual mass $m_r$) in the free
 domain wall propagator. This is useful in practice since
 we would like to apply the anisotropic lattice formalism
 to massive quarks (charm) and the extension in the fifth dimension
 is usually not very large in practical simulations.

 \subsection{The free propagator}
 \label{subsec:free_G}

 After performing the four-dimensional Fourier transform,
 the free domain wall fermion matrix appearing
 in action~(\ref{eq:fermion_action}) is given by:
 \ba
 \label{eq:D0}
 D^{(0)}_{ss'}(p) &=& \left[-b(p)+i\fslash{\tilde{p}}\right]\delta_{ss'}
 +\left[P_R\delta_{s+1,s'}+P_L\delta_{s-1,s'}\right]
 \nonumber \\
 &-& m\left(P_L\delta_{s,0}\delta_{s',L_s-1}
 +P_R\delta_{s',0}\delta_{s,L_s-1}\right)\;,
 \\
 \label{eq:D0_W}
 &=&\left[i\fslash{\tilde{p}}+W^{+}(p)\right]P_R
 +\left[i\fslash{\tilde{p}}+W^{-}(p)\right]P_L\;
 \ea
 where the function $b(p)$ and the notation $\fslash{\tilde{p}}$
 are defined as:
 \ba
 b(p) &=& 1-M_5+\sum_\mu\kappa_\mu(1-\cos p_\mu)\;,
 \nonumber \\
 \fslash{\tilde{p}}&=&\sum_\mu\gamma_\mu \tilde{p}_\mu
 \;,\;\;
 \tilde{p}_\mu=\kappa_\mu\sin p_\mu\;.
 \ea
 The matrices $W^{\pm}(p)$ appearing in Eq.~(\ref{eq:D0_W})
 are defined as:
 \be
 [W^+(p)]^\dagger_{ss'}\equiv\left[W^{-}(p)\right]_{ss'} = -b(p)\delta_{ss'} -m
 \delta_{L_s-1,s'}\delta_{s,0} +\delta_{s-1,s'}\;.
 \ee
 With this notation, one easily works out the second order
 operator $\Omega^{(0)}(p)$ which is defined as:
 \ba
 \label{eq:omegaLR}
 \Omega^{(0)}(p) &\equiv& D^{(0)}(p)\cdot\left[D^{(0)}(p)\right]^\dagger
 =\Omega^{(0)}_L(p)P_L+\Omega^{(0)}_R(p)P_R\;,
 \nonumber \\
 &=& \tilde{p}^2+W^+W^-P_R+W^-W^+P_L\;.
 \ea
 Note that the operators $\Omega^{(0)}_{L/R}(p)=\tilde{p}^2+W^\mp W^\pm$ have trivial
 Dirac structure and they are related to each other by:
 \be
 \label{eq:LRsymmetry}
 \left[\Omega^{(0)}_L(p)\right]_{s,s'}=
 \left[\Omega^{(0)}_R(p)\right]_{L_s-s-1,L_s-s'-1}\;.
 \ee
 The matrices $W^-W^+$ and $W^+W^-$ are referred to as
 the mass matrices for the right and left-handed fermions, respectively.
 Free fermion propagator is then expressed in terms of
 the inverse of the second order operators as:
 \ba
 \label{eq:SF}
 S_F(p) &\equiv & \left[D^{(0)}(p)]\right]^{-1}=
 \left[D^{(0)}(p)\right]^\dagger
 \cdot \left[\Omega^{(0)}(p)\right]^{-1}\;,
 \nonumber \\
 &=&\left[-i\fslash{\tilde{p}}+W^{-}(p)\right]P_RG^{(0)}_R
 +\left[-i\fslash{\tilde{p}}+W^{+}(p)\right]P_LG^{(0)}_L\;,
 \ea
 where we have used the notation:
 \be
 G^{(0)}_{L/R}(p)=\left[\Omega^{(0)}_{L/R}(p)\right]^{-1}\;.
 \ee
 The explicit results for $G^{(0)}_R(p)$ and $G^{(0)}_L(p)$ are
 found to be:
 \ba
 \label{eq:freeG}
 \left[G^{(0)}_R(p)\right]_{s,s'}&=&
 \left[G^{(\infty)}(p)\right]_{s,s'}
 +A_+e^{-\alpha_G(s+s')}+A_-e^{-\alpha_G(2L_s-2-s-s')}
 \nonumber\\
 &+&A_m(e^{-\alpha_G(L_s-1+s-s')}+e^{-\alpha_G(L_s-1+s'-s)})
 \;,\nonumber \\
 \left[G^{(0)}_L(p)\right]_{s,s'}&=&
 \left[G^{(\infty)}(p)\right]_{s,s'}
 +A_-e^{-\alpha_G(s+s')}+A_+e^{-\alpha_G(2L_s-2-s-s')}
 \nonumber\\
 &+&A_m(e^{-\alpha_G(L_s-1+s-s')}+e^{-\alpha_G(L_s-1+s'-s)})
 \;,
 \ea
 where $\left[G^{(\infty)}(p)\right]_{s,s'}$ is the corresponding
 Green's function for infinite fifth dimension, i.e. when the boundaries
 are absent. The other coefficients in this equation are all
 functions of the four-momentum $p$, the boundary mass parameter $m$ and the
 extension of the fifth dimension $L_s$. The explicit
 formulae are:
 \ba
 \label{eq:more_on_freeG}
 \left[G^{(\infty)}(p)\right]_{s,s'}&=&Be^{-\alpha_G\left| s-s'\right|}
 \;, \;\;\;B=\frac{1}{2b\sinh{\alpha_G}}\;,
 \nonumber \\
 A_{\pm}&=&\Delta^{-1}B(1-m^2)(e^{\mp\alpha_G}-b)\;,
 \nonumber \\
 A_m&=&-\Delta^{-1}m+\Delta^{-1}Bm_r\left(be^{-\alpha_G}-1
 +m^2(1-be^{\alpha_G})\right)\;,
 \nonumber \\
 \Delta&=&e^{2\alpha_G}(b-e^{-\alpha_G})\left(1+m^2m^2_{r}\right)
 +\left(m^2+m^2_{r}\right)(e^{\alpha_G}-b)
 \nonumber \\
 &+&2mm_{r}b(e^{2\alpha_G}-1)\;.
 \ea
 Here the (momentum-dependent) parameters $\alpha_G$ and $m_{r}$ are defined as:
 \be
 \label{eq:alphaG_def}
 \cosh{\alpha_G(p)}\equiv
 \frac{1+b^2(p)+\sum_{\mu}\kappa^2_{\mu}\sin^2{p_\mu}}{2b(p)}\;,
 \;\;
 m_r=e^{-\alpha_GL_s}\;,
 \ee

 In the perturbative calculation of fermion propagator,
 it is useful to choose a basis such that
 the mass matrix of the free fermion propagator is
 diagonal~\cite{aoki99:DWF_pert_selfenergy} in the
 fifth dimension. In other words,
 one seeks for unitary matrices $\calU^{(0)}$ and
 $\calV^{(0)}$ such that:
 \be
 \label{eq:diagonalization_free}
 \left[\calU^{(0)} W^-W^+\calU^{(0)\dagger}\right]_{ss'}=(M^2_0)_s\delta_{ss'}\;,
 \left[\calV^{(0)} W^+W^-\calV^{(0)\dagger}\right]_{ss'}=(M^2_0)_s\delta_{ss'}\;,
 \ee
 where $(M^2_0)_s$ for $s=0,1,\cdots,L_s-1$ are the corresponding eigenvalues.
 \footnote{The suffix $0$ in $(M^2_0)_s$ stands for free theory.}
 Note that due to the symmetry relation~(\ref{eq:LRsymmetry}),
 the spectrum of $W^-W^+$ and $W^+W^-$ are identical.
 However, the unitary transformation matrix $\calU^{(0)}$ is not
 the same as $\calV^{(0)}$, instead, they are related by:
 \be
 \calV^{(0)}_{ss'}=\calU^{(0)}_{s,L_s-1-s'}\;.
 \ee
 The unitary transformation matrices $\calU^{(0)}$ and $\calV^{(0)}$
 are obtained by finding all linearly independent eigen-modes corresponding to
 all possible eigenvalues $(M^2_0)_s$ in Eq.~(\ref{eq:diagonalization_free}).
 The properties of these eigen-modes can be easily obtained
 from corresponding formulae in  the isotropic case~\cite{neuberger98:residual_mass,aoki99:DWF_pert_selfenergy}.
 This set of basis is also useful when we discuss the
 renormalization factors of the chiral mode.

 \subsection{Dispersion relation of free domain wall fermions}
 \label{subsec:dispersion_free}

 By inspecting the low-momentum behavior of the
 free domain wall fermion propagator, one verifies that
 there exists a left-handed and a right-handed chiral pole in
 the corresponding fermion propagator near the two domain walls
 at $s=0$ and $s=L_s-1$, respectively. These chiral poles reflect
 the existence of the corresponding chiral mode bounded at these
 walls~\cite{shamir93:DWF_boundary},
 as can be easily verified by solving the corresponding
 Dirac equation for $D^{(0)}(p)$.
 The only complication that arises in the case of anisotropic lattice
 is that, in order to restore rotational symmetry for small
 lattice momenta, one has a relation among $\chi$, $\kappa_t$ and $\kappa_s$
 when other parameters (e.g. $M_5$, $m$ and $L_s$) are fixed.
 In the massless limit, i.e. both $m$ and $m_r$ being zero,
 this relation simplifies to: $\chi=\kappa_t/\kappa_s$.
 For massive fermions, however, the relation is quite complicated
 and we refer the reader to Appendix~\ref{asubsec:aniso} for
 the explicit formulae (c.f. Eq.~(\ref{eq:mQ}) and Eq.~(\ref{eq:chi})).
 Here, we will only outline the strategy to derive this relation
 and show some general features of it in figures.

 \begin{figure}[htb]
 \begin{center}
 \includegraphics[width=12.0cm,angle=0]{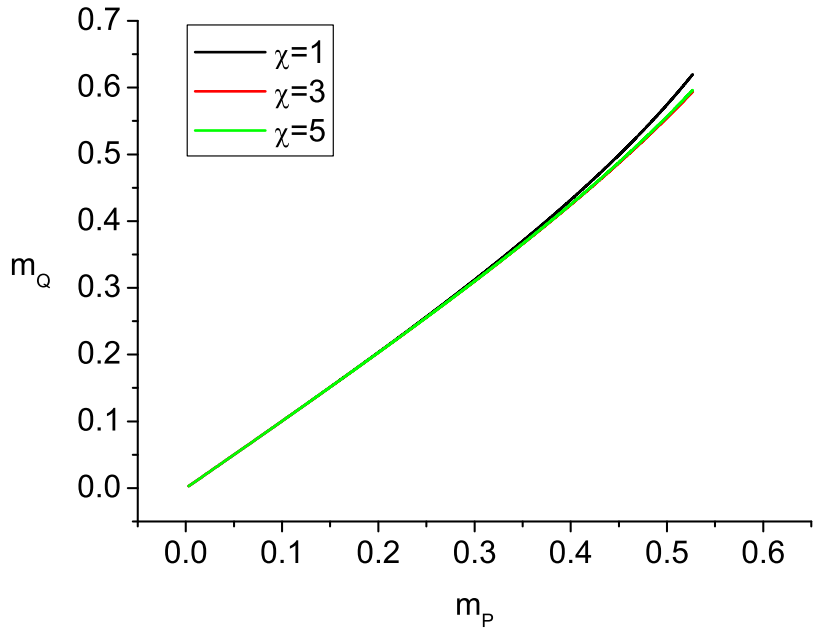}
 \end{center}
 \caption{The pole mass of the free domain wall quark, $m_Q$, measured
 in $1/a_s$ unit as a function of the propagator bare mass parameter $m_P=(1-b(0)^2)(m+m_r)$
 for three values of the anisotropy parameter $\chi$. Three curves
 corresponds to $\chi=1$, $3$ and $5$, respectively.
 We have set $\kappa_s=1$, $M_5=0.5$ and $L_s=8$ in this plot.}
 \label{fig:mQvsmP}
 \end{figure}
 \begin{figure}[htb]
 \begin{center}
 \includegraphics[width=12.0cm,angle=0]{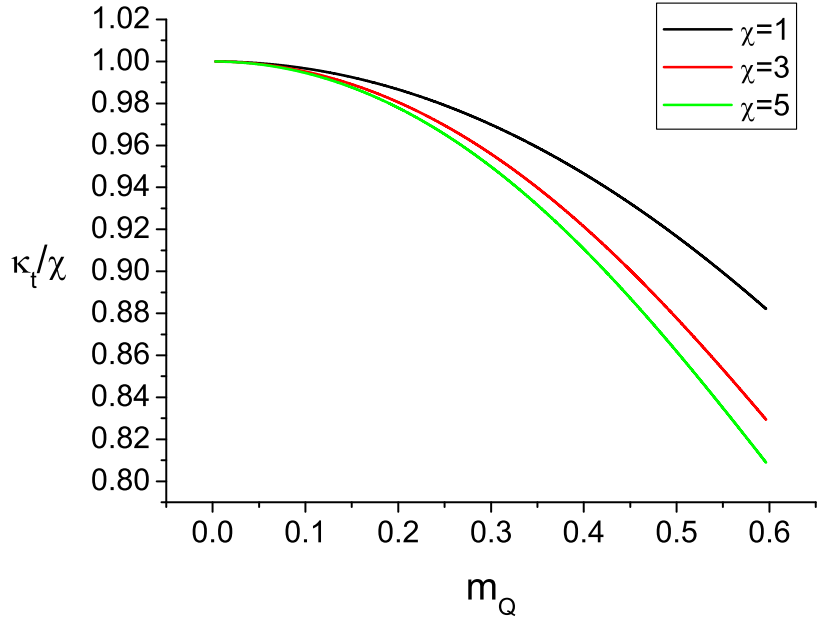}
 \end{center}
 \caption{With the same set of parameters as Fig.~\ref{fig:mQvsmP},
 the value of $\kappa_t/\chi$ is shown as
 a function of the pole mass  $m_Q$ for $\chi=1$, $3$ and $5$.
 It is seen that in the massless limit,
 the expected relation: $\chi=\kappa_t/\kappa_s$ is
 recovered for all $\chi$.}
 \label{fig:KtvsmQ}
 \end{figure}
 To obtain the dispersion relation for the free domain wall fermion,
 one searches for the pole of the free
 domain wall fermion propagator. The position of the pole in terms of
 four-momentum $p_\mu$ is given by: $p_\mu=(iE_\bp,\bp)$.
 This then gives the dispersion relation of the fermion: $E_\bp$
 as a function of the three-momentum $\bp$. The pole mass of
 the fermion is identified as: $m_Q\equiv E_{\bp=0}=E_0$.

 In Fig.~\ref{fig:mQvsmP}, we have illustrated
 the pole mass of
 the massive quark $m_Q$ (measured in $1/a_s$ unit)
 as a function of the bare propagator mass parameter $m_P$ defined as:
 \be
 \label{eq:mP_def}
 m_P\equiv (1-b(0)^2)(m+m_r)\;,
 \ee
 for three values of anisotropy $\chi$, namely $\chi=1$, $3$ and $5$,
 respectively. Note that the residue mass is given by
 $m_r=e^{-\alpha_G(im_Q,0,0,0)L_s}$, as explained in the appendix.
 The other relevant parameters in this plot are chosen as:
  $\kappa_s=1$, $M_5=0.5$ and $L_s=8$.
 It is seen that the pole mass $m_Q$ basically
 depends on $m_P$ linearly for $m_P$ values not too large.
 The nonlinearity only sets in for large bare quark mass values.
 Also, the dependence of $m_Q$ on $m_P$ saturates
 quickly for anisotropy larger than unity.
 For example, the difference between the case $\chi=3$ and $\chi=5$
 is barely visible.

 The dispersion relation for the free domain wall fermion
 is quite complicated. However, for small three lattice
 momenta: $a_s|\bp| \ll 1$, one finds that:
 \be
 E_\bp=m_Q + {\bp^2 \over 2M_{\rm kin}} + O (\bp^4)\;,
 \ee
 where $M_{\rm kin}$ is the so-called kinetic mass of the
 quark. To recover normal relativistic dispersion
 relation for small three lattice momenta, one requires:
 $m_Q=M_{\rm kin}$ which yields a relation
 among the pole mass and the other parameters in the theory
 (see Appendix~\ref{asubsec:aniso} for the explicit formulae).
 For the same set of bare parameters as in Fig.~\ref{fig:mQvsmP},
 we have shown the values of $\kappa_t/\chi$
 as a function of the pole mass $m_Q$ (measured in $1/a_s$ unit)
 in Fig.~\ref{fig:KtvsmQ}. Note that in a  quenched calculation,
 the anisotropy parameter $\chi$ is in fact fixed by the pure gauge sector.
 Therefore, once the pole mass of the quark (or equivalently the propagator
 mass $m_P$) and other bare
 parameters are fixed, one has to tune $\kappa_t$ in such a way that
 the dispersion relation of the fermion resembles the continuum
 form in the small $a_s|\bp|$ region.
 Basically one utilizes the relation shown in Fig~\ref{fig:KtvsmQ} to
 find out the appropriate value of $\kappa_t$ for a give value of $m_Q$.
 Of course, for vanishing $m_Q$, one recovers the expected relation:
 $\chi=\kappa_t/\kappa_s$.

 \section{Fermion Propagator to One-loop}

 In this section, we outline the basic strategies to calculate the domain wall fermion
 propagator to one-loop. Numerical results will be provided in the next section.
 To simplify the presentation here, some explicit formulae
 for the loop integrals are provided in Appendix~\ref{asec:longformulae}.

 \subsection{Feynman rules}
 \label{subsec:feynman_rule}

 First of all, one needs the free propagator of the
 lattice gauge fields~\cite{shigemitsu00:aniso_oneloop}.
 This is obtained by expressing the pure-gauge action~(\ref{eq:gauge_action})
 (with the tadpole improvement factor $u_s$ set to unity)
 into the gauge potential $A_\mu(x)$.
 One also needs to fix to a particular gauge.
 We adopt the following gauge-fixing action:
 \be
 \label{sgfa}
 S_{gf}= \frac{1}{2 \alpha_g} \, \frac{1}{\chi} \, \sum_x \left[
 \chi^2 \partial_t(a_t A_t) + \sum_j \partial_j(a_sA_j) \right]^2 ,
 \ee
 where $\alpha_g$ is the gauge-fixing parameter.
 The gauge field measure action and the Fadeev-Popov ghost action
 do not enter our one-loop calculation.

 After performing Fourier transformation of the gauge fields,
 the quadratic part of the gauge action then has the
 standard form in momentum space:
  \ba
 S^{(0)}_g[A_\mu] = \frac{1}{2} \, \sum_{\mu\nu}
 \int_{-\pi}^\pi \frac{d^4 l}{(2 \pi)^4} \left( \bar{A}_\mu(l)
 \,M_{\mu\nu}(l) \, \bar{A}_\nu(-l) \right) ,
 \ea
 where
 \ba
 M_{00} &=& \chi \, \left[ \frac{\chi^2}{\alpha_g} \hat{l}_0^2 \; + \;
 \sum_j
 \hat{l}_j^2 \, q_{0j} \right]   \\
 M_{jj} &=& \frac{1}{\chi} \, \left[ \frac{1}{\alpha_g} \hat{l}_j^2 \; +
 \; \chi^2 \, \hat{l}_0^2 \, q_{0j} \; + \; \sum_{j'\neq j}\hat{l}_{j'}^2 \,
 q_{j'j}
 \right]   \\
 M_{i \neq j} &=& \frac{1}{\chi} \, \left[ \frac{1}{\alpha_g}
 \hat{l}_i \hat{j}_j \; - \;  \hat{l}_i \hat{l}_j \, q_{ij} \right]   \\
 M_{0 j} &=& M_{j 0}  = \chi \, \left[ \frac{1}{\alpha_g}
 \hat{l}_0 \hat{l}_j \; - \;  \hat{l}_0 \hat{l}_j \, q_{0j} \right]
 \ea
 with lattice momentum defined as:
 $\hat{l}_\mu \equiv 2\,\sin(\frac{l_\mu}{2})$.
 The quantities $q_{\mu\nu}$ appearing in the above equations
 are given by:
 \be
 q_{ij} = 1 \; + \; \frac{1}{12} \, (\hat{l}^2_i + \hat{l}^2_j)
 \qquad i \neq j\;, \;\;
 q_{0j} = 1 \; + \; \frac{1}{12} \, \hat{l}^2_j\;.
 \ee
 Using these notations, the free gauge field propagator is expressed as:
 \ba
 \label{dmunu}
 D_{\mu\nu}(l) =
 M^{-1}_{\mu\nu}
 = \frac{1}{(\hat{l}^2)^2} \left[ \alpha_g \hat{l}_\mu
 \hat{l}_\nu \chi +
 \frac{f^{\mu\nu}(\hat{l}_\rho,q_{\rho\sigma},\chi)}
 {f_D(\hat{l}_\rho,q_{\rho\sigma},\chi)} \right] ,
 \ea
 For simplicity, we set $\alpha_g=1$ in the following calculation.
 The explicit expressions for $f^{\mu\nu}$ and  $f_D$ maybe found
 in the literature~\cite{shigemitsu00:aniso_oneloop}.

 The vertex functions can be obtained from the corresponding
 expression in the isotropic lattice case, which was given
 in Ref.~\cite{aoki99:DWF_pert_selfenergy}. After making necessary
 modifications to the case of anisotropic lattice, we have for
 the one gluon emission vertex:
 \be
 V_1(k,p,l;a,\mu)=-igT^a\kappa_\mu\left[\gamma_\mu\cos{1\over 2}(p_\mu-k_\mu)
 +i\sin{1\over 2}(p_\mu-k_\mu)\right]\;,
 \ee
 where $k$, $p$ and $l=-(k+p)$ are the momenta for
 the anti-quark, quark and the gluon field, respectively;
 $a$ and $\mu$ representing the color and Lorentz index
 of the gluon. The anti-quark, quark and two-gluon
 vertex is given by:
 \ba
 V_2(k,p,l_1,l_2;a,b,\mu,\nu) &=& {g^2\over 2}{\kappa_\mu\over 2}\{T^a,T^b\}
 \left[i\gamma_\mu\sin{1\over 2}(p_\mu-k_\mu)\right.
 \nonumber \\
 &+&\left.\cos{1\over 2}(p_\mu-k_\mu)\right]\delta_{\mu\nu}\;,
 \ea
 where $k$, $p$, $l_1$ and $l_2$ are the momenta for
 anti-quark, quark and the two gluon fields with
 the energy-momentum conservation constraint: $k+p+l_1+l_2=0$.
 $\mu$, $\nu$, $a$ and $b$ are the Lorentz and color indices
 for the two gluons. Expressions for other vertices will not
 be needed in our one-loop calculation.

 \subsection{Fermion self-energy}
 \label{subsec:self_energy}

 To one-loop order, two Feynman diagrams contribute to fermion
 self-energy: the tadpole diagram and the half-circle diagram.
 Thus, the 1PI fermion 2-point vertex function reads:
 \be
 V^{(2)}_{\rm 1-loop} (p)_{s,s'} =
 \left[ i\sum_{\mu}\kappa_{\mu} \gamma_\mu \sin p_\mu +
 W^+ (p) P_R + W^- (p) P_L-\Sigma (p) \right]_{s,s'}
 \ee
 where the fermion self-energy $\Sigma (p)$ receives contributions from
 the tadpole and the half-circle diagrams, respectively:
 \be
 \Sigma (p) = \Sigma^{\rm tadpole} (p) + \Sigma^{\rm half-circle} (p).
 \ee

 Using the Feynman rules given in the previous section,
 the contribution from the tadpole diagram can be written as:
 \ba
 \label{eq:tadpole}
 \Sigma^{\rm tadpole}(p) &=& \frac{1}{2} g^2 C_F \sum_\mu \kappa_\mu (i \gamma_\mu
 \sin p_\mu + \cos p_\mu )\nonumber \\
 &\times &\int_{-\pi}^\pi \frac{d^4 l}{(2 \pi)^4}
 \frac{1}{(\chi^2\hat{l}^2_0+\sum_j \hat{l}^2_j)^2}
 (\hat{l}^2_\mu\chi+\frac{f^{\mu\mu}}{f_D}) \delta_{s,s'}
 \;,
 \nonumber \\
 &=&-\left[\sum_\mu i\gamma_\mu \tilde{p}_\mu I_{\rm tad,\mu}(s,s')+
 M_{\rm tad}(s,s')\right]
 \ea
 where the explicit expressions for
 $I_{\rm tad,\mu}(s,s')$ and $M_{\rm tad}(s,s')$ are given
 by Eq.~(\ref{eq:IMtad}) in appendix~\ref{asec:longformulae}.
 Note that both $I_{\rm tad,\mu} (s, s')$ and $M_{\rm tad}(s,s')$ are
 independent of the quark mass explicitly.
 They are diagonal in the fifth dimensional index and
 can be computed by numerical integration.

 For the half circle diagram, the contribution can be written as:
 \ba
 \label{eq:half}
 \Sigma^{\rm half-circle}_{s,s'} (p)&=& \int_{-\pi}^\pi
 \frac{d^4 l}{(2 \pi)^4} \sum_\mu (-g^2C_F)\kappa^2_\mu
 \left\{ \gamma_\mu \cos (\frac{l + p}{2})_\mu
  + i \sin (\frac{l + p}{2})_\mu \right\}\nonumber \\
 &\times& S_{F} (l)_{s,s'} \times  \left\{ \gamma_\mu \cos (\frac{l + p}{2})_\mu
  + i \sin (\frac{l + p}{2})_\mu \right\}\nonumber \\
   &\times& \left\{ \frac{1}{(\chi^2(\widehat{p_0-l_0})^2+\sum_j(\widehat{p_j-l_j})^2)^2}
  \left[\chi(\widehat{p_\mu-l_\mu})^2+\frac{f^{\mu\mu}}{f_D}\right] \right\}
  \;.
 \ea
 Note that although $\Sigma^{\rm tadpole}(p)$ is proportional to the unit matrix
 in the fifth dimensional space, $\Sigma^{\rm half-circle}_{s,s'}(p)$ is
 non-trivial due to the fermion propagator $S_F(l)_{ss'}$ which
 is given by Eq.~(\ref{eq:SF}) in the previous section.
 To simplify our one-loop calculation, we have neglected the effects
 caused by the finite extension of the fifth dimension in the following
 calculation, i.e.
 we have set: $m_r=e^{-\alpha_GL_s}\simeq 0$ in the quark
 propagator $S_F(l)_{ss'}$.
 However, in order to study the quark mass effects,
 we have kept the bare quark mass parameter $m$ non-zero.

 The calculation of $\Sigma^{\rm half-circle}_{s,s'}(p)$ is
 somewhat different in the massive and the massless cases.
 If the fermion is massless,
 it is well-known that $\Sigma^{\rm half-circle}_{s,s'}(p)$ contains
 infra-red divergences~\cite{shigemitsu00:aniso_oneloop,aoki99:DWF_pert_selfenergy}.
 If the fermion is massive, no infra-red divergence shows up in this contribution.
 In the massive fermion case, since there is no
 infra-red divergence, the expression given
 in Eq.~(\ref{eq:half}) can be evaluated directly
 by numerical integration methods. In the massless case, however,
 the infra-red divergent part (which can be computed analytically)
 has to be subtracted from the self-energy contribution
 $\Sigma^{\rm half-circle}_{s,s'}(p)$
 before it can be evaluated numerically.
 For the physical quantities in the massive fermion case,
 we denote them as $f(m)$. For the same quantity in the massless case,
 we denote the subtracted expression as:
 \be
 \label{eq:subtract_def}
 f_{\rm finite}=f(m=0)-f_{\log}\;,
 \ee
 where $f$ represents some physical quantity and
 $f_{\log}$ contains the logarithmical infra-red divergent
 part to be removed from $f(m=0)$. Usually, $f_{\log}$ is taken
 to be the corresponding infra-red divergent contribution in
 the continuum which can be computed analytically. After the
 subtraction of infra-red divergent part, $f_{\rm finite}$
 is free of infra-red divergences and can be evaluated numerically.

 For the half-circle contribution in the massive case,
 we obtain:
 \ba
 \label{eq:halfcircle}
 \Sigma^{\rm half-circle}_{s,s'}(m)& = &- i \sum_\mu\gamma_\mu\tilde{p}_\mu
 \left( [I^+_\mu(m)]_{s,s'} P_R + [I^-_\mu(m)]_{s, s'} P_L \right) \nonumber \\
 & &- \left([M^+(m)]_{s, s'} P_R + [M^-(m)]_{s, s'}
 P_L\right)\;.
 \ea
 In the massless case, after separating the
 infra-red divergent part, the half-circle
 contribution are obtained as:
 \ba
 \label{eq:halfcircle_massless}
 \Sigma^{\rm half-circle}_{s,s'}& = &- i \sum_\mu\gamma_\mu\tilde{p}_\mu
 \left[ \left(I_{\rm finite,\mu}^+ (s,s')+I_{\rm log,\mu}^+ (s,s')\right)P_R \right.
 \nonumber \\
 & &+\left.\left(I_{\rm finite,\mu}^-(s, s')+I_{\rm log,\mu}^- (s,s') \right)P_L \right]
 \nonumber \\
 & &- \left[\left(M_{\rm finite}^+(s, s')+M_{\rm log}^+(s,s')\right) P_R \right.
 \nonumber \\
 & &+ \left.(\left(M_{\rm finite}^-(s, s')+M_{\rm log}^-(s,s')\right)
 P_L \right]
 \ea
 where explicit formulae for the quantities appearing in
 the above formulae can be found in appendix B.

 Combining the tadpole and the half-circle contributions,
 we may express the effective action of the domain wall fermion as:
 \ba
 \Gamma^{(2)} = \bar{\psi}(-p)_s \left[\sum_\mu
  i \gamma_\mu \tilde{p}_\mu \left( Z^+_\mu P_R + Z^-_\mu P_L \right)
 + \overline{W}^+ P_R + \overline{W}^- P_L \right]_{s , s'} \psi (p)_s'
 \ea
 where the matrices $Z^\pm_\mu$ and $\bar{W}^\pm$ are given by:
 \ba
  \label{eq:z-factor}
 Z^\pm_\mu(s, s') &=& \delta_{s, s'} + I_{\rm tad,\mu}(s, s')
       + [I^\pm_\mu(m)]_{s, s'} \;,
 \nonumber
 \\
 \overline{W}^\pm(s, s') &=& W^\pm(0)(s, s') + M_{\rm tad}(s, s')
       + [M^\pm(m)]_{s, s'} \;.
 \ea
 in the massive case. The quantities
 $[I^\pm_\mu(m)]_{s, s'}$ and $[M^\pm(m)]_{s, s'}$ are replaced
 by the sum of a finite part and an infra-red divergent part
 in the massless case:
 \ba
  \label{eq:z-factor0}
 Z^\pm_\mu(s, s') &=& \delta_{s, s'} + I_{\rm tad,\mu}(s, s')
       + I^\pm_{\log,\mu}(s,s')+I^\pm_{\rm finite,\mu}(s,s') \;,
 \nonumber
 \\
 \overline{W}^\pm(s, s') &=& W^\pm(0)(s, s') + M_{\rm tad}(s, s')
       + M^\pm_{\rm log}(s,s')+M^\pm_{\rm finite}(s,s') \;.
 \ea
 The explicit formulae for the tadpole contributions $I_{\rm tad,\mu}(s,s')$,
 $M_{\rm tad}(s,s')$ and the half-circle contributions
 $[I^\pm_\mu(m)]_{s, s'}$, $[M^\pm(m)]_{s, s'}$
 $M^\pm_{\rm log}(s,s')$, $M^\pm_{\rm finite}(s,s')$,
 $I^\pm_{\log,\mu}(s,s')$, $I^\pm_{\rm finite,\mu}(s,s')$ can be found
 in appendix~\ref{asec:longformulae}.

 \subsection{Wave-function and mass renormalization of the chiral mode}
 \label{subsec:Z}

 We are concerned with the renormalization of the chiral modes
 which are bound to the walls. To obtain this information,
 it is better to use a new basis, $\psi^d(p)$,
 which diagonalizes the mass matrices $\overline{W}^{\pm}$
 in Eq.~(\ref{eq:z-factor}).
 This new basis is related to the old one via:
 \ba
 \psi^d_s (p) = \calU_{s,s'}P_R\psi_{s'}(p) + \calV_{s,s'} P_L\psi_{s'}(p),
 \ea
 where the two unitary matrices $\calU$ and $\calV$ satisfy
 \be
 \left[ \calU \overline{W}^-\overline{W}^+ \calU^\dagger \right]_{s,s'} = M_s^2
 \delta_{s,s'}\;,\;\;
 \left[ \calV \overline{W}^+\overline{W}^- \calV^\dagger \right]_{s,s'} = M_s^2
 \delta_{s,s'} \;.
 \ee
 Under this new basis, the effective action of the fermion becomes:
 \ba
 \Gamma^{(2)} &=& \bar{\psi}^d (-p)_s \Bigg[\sum_\mu
  i \gamma_\mu \tilde{p}_\mu \left(\calU Z^+_\mu \calU^\dagger P_R
  +\calV Z^-_\mu \calV^\dagger P_L
  \right)\nonumber \\
 &+&\calV \overline{W}^+ \calU^\dagger P_R
 +\calU \overline{W}^- \calV^\dagger P_L \Bigg]_{s , s'}
 \psi^d (p)_{s'}\;.
 \ea
 The unitary matrices $\calU$, $\calV$ and the corresponding
 eigenvalues $(M^2)_s$ can be calculated to 1-loop level:
 \ba
 \calU &=& (1 + g^2 \calU^{(1)})\calU^{(0)}\;,
 \;\; \calV=(1+g^2 \calV^{(1)}) \calV^{(0)}\;,
 \\
 (M^2)_s &=& (M^2_0)_s+g^2(M^2_1)_s\;.
 \ea
 where tree-level matrices $\calU_0$ and $\calV_0$ are
 defined in Eq.~(\ref{eq:diagonalization_free}).

 After diagonalization of the mass matrix,
 the effective action for
 the chiral mode field $\psi^d(p)_{L_s-1} = \chi_0(p)$ becomes:
 \ba
 \bar\chi_0(-p) \left[\sum_\mu i \gamma_\mu \tilde{p}_\mu
 \left( \tilde{Z}_{+,\mu} P_R +
 \tilde{Z}_{-,\mu} P_L \right)+\tilde{W}_+ P_R +
 \tilde{W}_- P_L \right] \chi_0(p)\;,
 \ea
 where
 \ba
 \label{eq:tildeZ}
 \tilde{Z}_{\pm,\mu} &=& 1+g^2 \left(I^d_{\pm, \rm tad,\mu}
 + I^d_{\pm,\mu}(m) \right)_{L_s-1,L_s-1}\;,
 \\
 \label{eq:tildeW}
 \tilde{W}_\pm &=& -m_P \left[ 1+g^2 \left(M^d_{\pm, \rm tad}
 + M^d_\pm(m) \right)_{L_s-1,L_s-1} \right] \;.
 \ea
 The above formulae work for the massive case. In the massless
 case, one has to modify the right-hand side of
 Eq.~(\ref{eq:tildeZ}) and Eq.~(\ref{eq:tildeW}) accordingly
 as specified by Eq.~(\ref{eq:subtract_def}).
 Note that chiral symmetry ensures that the effective quark
 mass term is renormalized by a multiplicative factor, i.e. there
 is no additive renormalization.
 The matrix elements of various $I^d_{\mu}$ and $M^d$ in the above
 expressions are given explicitly by:
 \ba
 g^2 I^d_{+,\rm tad,\mu} &=& \calU^{(0)} I_{\rm tad,\mu} \calU^{(0)\dagger} \;,
 \;\;
 g^2 I^d_{-,\rm tad,\mu} = \calV^{(0)} I_{\rm tad,\mu} \calV^{(0)\dagger} \;,
 \nonumber
 \\
 g^2 I^d_{+,\mu} &=& \calU^{(0)} I_{\mu}^+ \calU^{(0)\dagger} \;,
 \;\;
 g^2 I^d_{-,\mu} = \calV^{(0)} I_{\mu}^- \calV^{(0)\dagger}\;,
 \nonumber
 \\
 g^2 I^d_{+,\rm log,\mu} &=& \calU^{(0)} I_{\rm log,\mu}^+ \calU^{(0)\dagger} \;,
 \;\;
 g^2 I^d_{-,\rm log,\mu} = \calV^{(0)} I_{\rm log,\mu}^- \calV^{(0)\dagger}\;,
 \nonumber
 \\
 g^2 I^d_{+,\rm finite,\mu} &=& \calU^{(0)} I_{\rm finite,\mu}^+ \calU^{(0)\dagger} \;,
 \;\;
 g^2 I^d_{-,\rm finite,\mu} = \calV^{(0)} I_{\rm finite,\mu}^- \calV^{(0)\dagger}\;.
 \\
 g^2m_P M^d_{+,\rm tad} &=& \calV^{(0)} M_{\rm tad} \calU^{(0)\dagger}\;,
 \;\;
 g^2m_P M^d_{-,\rm tad} = \calU^{(0)} M_{\rm tad} \calV^{(0)\dagger}\;,
 \nonumber
 \\
 g^2m_P M^d_+ &=& \calV^{(0)} M^+ \calU^{(0)\dagger}\;,
 \;\;
 g^2m_P M^d_- = \calU^{(0)} M^- \calV^{(0)\dagger}\;,
 \nonumber
 \\
 g^2m_P M^d_{+,\rm log} &=& \calV^{(0)} M_{\rm log}^+ \calU^{(0)\dagger}\;,
 \;\;
 g^2m_P M^d_{-,\rm log} = \calU^{(0)} M_{\rm log}^- \calV^{(0)\dagger}\;,
 \nonumber
 \\
 g^2m_P M^d_{+,\rm finite} &=& \calV^{(0)} M_{\rm finite}^+ \calU^{(0)\dagger}\;,
 \;\;
 g^2m_P M^d_{-,\rm finite} = \calU^{(0)} M_{\rm finite}^- \calV^{(0)\dagger}\;.
 \ea
 It is also easy to show that:
 \ba
 \label{eq:Idtad&Idlog_def}
 (I^d_{+,\rm tad,\mu})_{L_s-1,L_s-1} &=& (I^d_{-,\rm tad,\mu})_{L_s-1,L_s-1}
 \equiv I^d_{\rm tad,\mu} \;,\nonumber
 \\
 (I^d_{+,\mu})_{L_s-1,L_s-1} &=& (I^d_{-,\mu})_{L_s-1,L_s-1}
 \equiv I^d_\mu \;,\nonumber
 \\
 (I^d_{+,\rm log,\mu})_{L_s-1,L_s-1} &=& (I^d_{-,\rm log,\mu})_{L_s-1,L_s-1}
 \equiv I^d_{\rm log,\mu} \;,\nonumber
 \\
 (I^d_{+,\rm finite,\mu})_{L_s-1,L_s-1} &=& (I^d_{-,\rm finite,\mu})_{L_s-1,L_s-1}
 \equiv I^d_{\rm finite,\mu} \;,\nonumber
 \\
 (M^d_{+,\rm tad})_{L_s-1,L_s-1} &=& (M^d_{-,\rm tad})_{L_s-1,L_s-1}
 \equiv M^d_{\rm tad}\;,\nonumber
 \\
 (M^d_+)_{L_s-1,L_s-1} &=& (M^d_-)_{L_s-1,L_s-1}
 \equiv M^d\;,\nonumber
 \\
 (M^d_{+,\rm log})_{L_s-1,L_s-1} &=& (M^d_{-,\rm log})_{L_s-1,L_s-1}
 \equiv M^d_{\rm log}\;,\nonumber
 \\
 (M^d_{+,\rm finite})_{L_s-1,L_s-1} &=& (M^d_{-,\rm finite})_{L_s-1,L_s-1}
 \equiv M^d_{\rm finite}\;,
 \ea
 Thus one obtains: $\tilde{Z}_{+,\mu} =\tilde{Z}_{-,\mu} \equiv
  \tilde{Z}_\mu$ and  $\tilde{W}_+ =\tilde{W}_- \equiv \tilde{W}$.
 Using the above notations, we get for the wave-function and
 mass renormalization factors:
 \ba
 \label{eq:tildeZ2}
 \tilde{Z}_{\mu} &=& 1+g^2 \left(I^d_{\rm tad,\mu} + I^d_{\mu}(m)\right) \;,
 \\
 \label{eq:tildeW2}
 \tilde{W}_\pm &=& -m_P \left[ 1+g^2 \left(M^d_{\rm tad}
 + M^d(m) \right)\right] \;,
 \ea
 in the massive case and
 \ba
 \label{eq:tildeZ3}
 \tilde{Z}_{\mu} &=& 1+g^2 \left(I^d_{\rm tad,\mu}
 +I^d_{\log,\mu}+I^d_{\rm finite,\mu}\right) \;,
 \\
 \label{eq:tildeW3}
 \tilde{W}_\pm &=& -m_P \left[1+g^2 \left(M^d_{\rm tad}
 + M^d_{\log}+M^d_{\rm finite} \right)\right] \;,
 \ea
 in the massless case.

 The tadpole contributions $I^d_{\rm tad,\mu}$ and $M^d_{\rm tad}$
 can be obtained numerically by using the
 definition~(\ref{eq:Idtad&Idlog_def}) and
 the explicit expressions of $I_{\rm tad,\mu}(s,s')$ and
 $M_{\rm tad}(s,s')$ given in
 appendix~\ref{asec:longformulae} (c.f. Eq.~(\ref{eq:IMtad})).
 Using the definition of $\calU^{(0)}$ and $\calV^{(0)}$, we get:
 \ba
 \label{eq:Idtadmu}
 I^d_{\rm tad,\mu}= - C_F \frac{1}{2}\int_{-\pi}^\pi \frac{d^4 l}{(2 \pi)^4}
 \frac{1}{(\chi^2\hat{l}^2_0+\sum_j \hat{l}^2_j)^2}
 (\hat{l}^2_\mu\chi+\frac{f^{\mu\mu}}{f_D})\;.
 \ea
 and if we define $M^0_{\rm tad}$ as
 \ba
 M^0_{\rm tad}=- C_F \frac{1}{2}\int_{-\pi}^\pi \frac{d^4 l}{(2 \pi)^4}
 \sum_\mu\frac{\kappa_\mu}{(\chi^2\hat{l}^2_0+\sum_j \hat{l}^2_j)^2}
 (\hat{l}^2_\mu\chi+\frac{f^{\mu\mu}}{f_D})\;,
 \ea
 we arrive at:
 \be
 \label{eq:Mdtad}
 M^d_{\rm tad}=2e^{-\alpha}/((1-b^2(0))(1+m^2e^{-2\alpha}))M^0_{\rm tad}
 \;,
 \ee
 where the parameter $\alpha$ is exponential decaying
 rate for the chiral mode in the fifth dimension.
 Using these equations, numerical values of $I^d_{\rm tad,\mu}$
 and $M^d_{\rm tad}$ will be presented in the next section.

 We now proceed to evaluate  the half-circle
 contributions: $ I^d_\mu$ and $M^d$.
 If we use the shorthand notation:
 \[
 \langle F\rangle_\calU \equiv
 \sum_{s,s'}(\calU^{(0)})_{L_s-1,s} F(s,s')
 (\calU^{(0)})_{L_s-1,s'}
 \]
 and a similarly notation associated with $\calV^{(0)}$,
 we have:
 \ba
 \langle G_L\rangle_\calU & = & B \left[
 \frac{\sinh\alpha_G -\sinh\alpha}{\cosh\alpha_G
 -\cosh\alpha}
 -(1-e^{-2\alpha})\frac{\calN_1}{\calD}
 \right]=\langle G_R\rangle_\calV \equiv \widetilde{G_L}\;,
 \nonumber \\
 \langle G_R\rangle_\calU & = & B\left[
 \frac{\sinh\alpha_G -\sinh\alpha}{\cosh\alpha_G
 -\cosh\alpha}
 -(1-e^{-2\alpha}) \frac{\calN_2}{\calD}
 \right]=\langle G_L\rangle_\calV \equiv\widetilde{G_R}\;,
 \nonumber \\
 \langle W_0^+ G_L \rangle_\calU
 &=& \langle W_0^- G_R \rangle_\calV =
 (e^{-\alpha}-b)\widetilde{G_R}+R_0+R_1\;,
 \nonumber \\
 \langle W_0^+ G_L\rangle_\calV
 &=& \langle W_0^- G_R\rangle_\calU =
 (e^{-\alpha}-b)\widetilde{G_R}+R_0+R_2 \;,
 \nonumber
 \ea
 where:
 \ba
 \calN_1&=&(1-m^2)(e^{\alpha_G}-b)+(1-m^2)(e^{-\alpha_G}-b)m^2e^{-2\alpha}
 +4bm^2\sinh \alpha_G e^{-\alpha}\nonumber \\
 \calN_2&=&(1-m^2)(e^{-\alpha_G}-b)+(1-m^2)(e^{\alpha_G}-b)m^2e^{-2\alpha}
 +4bm^2\sinh \alpha_G e^{-\alpha}\nonumber \\
 \calD &=&(1+m^2e^{-2\alpha})\left[(e^{-\alpha_G}-b)
 +m^2(be^{-2\alpha_G}-e^{-\alpha_G})\right](e^{\alpha_G}-e^{-\alpha})^2
 \nonumber \\
 R_0&=&\frac{1-e^{-2\alpha}}{1+m^2e^{-2\alpha}} \frac{1}{(1-e^{-(\alpha+\alpha_G)})^2}
 \left\{Bm^2(1-e^{-2\alpha_G})e^{-3\alpha}\right.\nonumber \\
 &-&\left. A_m me^{\alpha_G}(e^{-\alpha_G}+2e^{-\alpha}-2e^{-\alpha_G}e^{-2\alpha}
 -e^{-3\alpha})\right\} \nonumber \\
 R_1&=&-A_m m\frac{1-e^{-2\alpha}}{1+m^2e^{-2\alpha}} \frac{m^2e^{-2\alpha}}
 {1-e^{-(\alpha+\alpha_G)}} \nonumber \\
 R_2&=&-A_m m\frac{1-e^{-2\alpha}}{1+m^2e^{-2\alpha}} \frac{1}
 {1-e^{-(\alpha+\alpha_G)}} \nonumber
 \ea
 Using the above notations, after some algebra, $I^d_\mu$ is found to be:
 \ba
 \label{eq:Idmu}
 I^d_\mu(m) &=& C_F \int \frac{d^4 l}{(2\pi)^4}
 \Biggl\{\Biggl[ \kappa_\mu\left( \cos l_\mu (
 (e^{-\alpha}-b)\widetilde{G_R}+R_0+\frac{R_1+R_2}{2})\right. \nonumber
 \\
 &+&\left.\frac{1}{2} \kappa_\mu\sin^2l_\mu (\widetilde{G_L}+\widetilde{G_R})
 \right)\Biggr]\frac{\chi\hat{l}^2_\mu+
 \frac{f^{\mu\mu}(l)}{f_D(l)}}{(\chi^2\hat{l}^2_0+\sum_j\hat{l}^2_j)^2}\nonumber
 \\
 &+&\Biggl[\left(\kappa_\mu
 ( (e^{-\alpha}-b)\widetilde{G_R}+R_0+\frac{R_1+R_2}{2})-2\kappa^2_\mu\cos^2 l_\mu/2
 \widetilde{G_L}\right)\nonumber \\
 &\times&\left(\frac{4(\chi^2)^{\delta_{\mu0}}}{\chi^2\hat{l}^2_0+\sum_j\hat{l}^2_j}
 (\chi\hat{l}^2_\mu+\frac{f^{\mu\mu}(l)}{f_D(l)})-2\chi
 -\frac{f^\mu_\mu(l)}{f_D(l)}+\frac{f^D_\mu(l)f^{\mu\mu}(l)}{f^2_D(l)}\right)\nonumber
 \\
 &-&2\chi\kappa^2_\mu
 \left(\sin^2 l_\mu/2 \widetilde{G_R}+\cos^2 l_\mu/2 \widetilde{G_L}\right)\nonumber
 \\
 &+&\sum_\nu \kappa^2_\nu
 \left(\sin^2 l_\nu/2 \widetilde{G_R}+\cos^2 l_\nu/2 \widetilde{G_L}\right) \nonumber
 \\
 &\times&\left(\frac{4(\chi^2)^{\delta_{\mu0}}}{\chi^2\hat{l}^2_0+\sum_j\hat{l}^2_j}
 (\chi\hat{l}^2_\nu+\frac{f^{\nu\nu}(l)}{f_D(l)})-\frac{f^\nu_\mu(l)}{f_D(l)}+
 \frac{f^D_\mu(l)f^{\nu\nu}(l)}{f^2_D(l)}\right)\Biggr]\nonumber
 \\
 &\times&\frac{\sin^2
 l_\mu}{(\chi^2\hat{l}^2_0+\sum_j\hat{l}^2_j)^2}\Biggr\}
 \ea

 In the massless case, what needs to be calculated numerically is
 the subtracted finite part $I^d_{\rm finite,\mu}$ defined as:
 \ba
 \label{eq:Idmu_finite}
 I^d_{\rm finite,\mu}&=&I^d_\mu(m=0) -C_F
 \frac{(1-b^2(0))(1-e^{-2\alpha})}{(1-b(0)e^{-\alpha})^2}
 \nonumber \\
 &\times&\int \frac{d^4 l}{(2\pi)^4}\frac{4\chi(\chi^2)^{\delta_{\mu0}}l^2_\mu\theta
 (\pi^2-(\chi^2l^2_0+\sum_j l^2_j))}
 {(\chi^2l^2_0+\sum_j l^2_j)^3}\;.
 \ea
 The infra-red divergent part that is subtracted from
 $I^d_\mu(0)$ is denoted as $I^d_{\log,\mu}$ and
 can be obtained analytically by using
 Eq.~(\ref{eq:Ilog}) and the
 definition~(\ref{eq:Idtad&Idlog_def}).

 For the contribution $M^d$, we have:
 \ba
 (\calV^{(0)}0 G_L \calU^{(0)\dagger})_{L_s-1,L_s-1}
 &=&(\calU^{(0)} G_R \calV^{(0)\dagger})_{L_s-1,L_s-1}=\widetilde{G^M_L}
 \;,\nonumber \\
 (\calV^{(0)} G_R \calU^{(0)\dagger})_{L_s-1,L_s-1}
 &=&(\calU^{(0)} G_R \calV^{(0)\dagger})_{L_s-1,L_s-1}=\widetilde{G^M_R}
 \;.
 \ea
 One also verifies that:
 \ba
 \widetilde{G^M_L} &=& \widetilde{G^M_R}
 \equiv-m_P\widetilde{G^M}\nonumber \\
 &=&-m_P\frac{(1-e^{-2\alpha})B}{(1-b^2(0))(1+m^2e^{-2\alpha})}
 \left[2\frac{1+e^{-(\alpha_G+\alpha)}}{1-e^{-(\alpha_G+\alpha)}}
 \frac{e^{-\alpha}}{1-e^{-2\alpha}}\right.
 \nonumber \\
 &+&\left.\frac{\Delta^{-1}}{(1-e^{-(\alpha_G+\alpha)})^2}\left(
 2e^{-\alpha}(1-m^2)(\cosh \alpha_G -b)+\frac{1+m^2e^{-2\alpha}}{B}\right)\right]
 \;.
 \ea
 We also find:
 \ba
 (\calV^{(0)} W^+G_L \calU^{(0)\dagger})_{L_s-1,L_s-1}
 &=&(\calU^{(0)} W^-G_R \calV^{(0)\dagger})_{L_s-1,L_s-1}\nonumber \\
 &=&-m_P\left((e^{-\alpha}-b)\widetilde{G^M}+R_0^M+R_1^M\right)\;,
 \nonumber \\
 (\calV^{(0)} W^-G_R \calU^{(0)\dagger})_{L_s-1,L_s-1}
 &=&(\calU^{(0)} W^+G_L \calV^{(0)\dagger})_{L_s-1,L_s-1}\nonumber \\
 &=&-m_P\left((e^{-\alpha}-b)\widetilde{G^M}+R_0^M+R_2^M\right)
 \;, \nonumber
 \\
 (\calV^{(0)} W^+(0) \calU^{(0)\dagger})_{L_s-1,L_s-1}
 &=&(\calU^{(0)} W^-(0)
 \calV^{(0)\dagger})_{L_s-1,L_s-1}\nonumber \\
 &=&-m_P\frac{1-e^{-2\alpha}}{(1-b^2(0))(1+m^2e^{-2\alpha})}\nonumber \\
 &\times&\left(1+\frac{2e^{-\alpha}(e^{-\alpha}-b(0))}{1-e^{-2\alpha}}\right)
 \;,\nonumber
 \ea
 where
 \ba
 R_0^M&=&\frac{(1-e^{-2\alpha})B\Delta^{-1}}{(1-b^2(0))(1+m^2e^{-2\alpha})}
 \frac{e^{-\alpha}}{(1-e^{-(\alpha_G+\alpha)})^2}\nonumber
 \\
 &\times&(1-m^2)\left[(e^{-\alpha_G}-e^{-\alpha})(e^{\alpha_G}-b)
 +(e^{\alpha_G}-e^{-\alpha})(e^{-\alpha_G}-b)\right]
 \nonumber \\
 R_1^M&=&\frac{1-e^{-2\alpha}}{(1-b^2(0))(1+m^2e^{-2\alpha})}
 \nonumber \\
 &\times&\left\{B\frac{1+2e^{-(\alpha_G+\alpha)}}{1-e^{-(\alpha_G+\alpha)}}
 +{1\over\Delta}\left(B\frac{(1-m^2)(e^{\alpha_G}-b)}{1-e^{-(\alpha_G+\alpha)}}
 \right.\right. \nonumber \\
 &+&\left.\left.\frac{(e^{-\alpha_G}-e^{-\alpha})m^2e^{-2\alpha}+e^{\alpha_G}-e^{-\alpha}}
 {(1-e^{-(\alpha_G+\alpha)})^2}+\frac{m^2e^{-\alpha}}{1-e^{-(\alpha_G+\alpha)}}\right)\right\}\;,
 \nonumber \\
 R_2^M&=&\frac{1-e^{-2\alpha}}{(1-b^2(0))(1+m^2e^{-2\alpha})}
 \nonumber \\
 &\times&\left\{B\frac{m^2e^{-2\alpha}+2e^{-(\alpha_G+\alpha)}}{1-e^{-(\alpha_G+\alpha)}}
 +{1\over\Delta}\left(B\frac{(1-m^2)(e^{\alpha_G}-b)}{1-e^{-(\alpha_G+\alpha)}}m^2e^{-2\alpha}
 \right.\right. \nonumber \\
 &+&\left.\left.\frac{(e^{\alpha_G}-e^{-\alpha})m^2e^{-2\alpha}+e^{-\alpha_G}-e^{-\alpha}}
 {(1-e^{-(\alpha_G+\alpha)})^2}+\frac{m^2e^{-\alpha}}{1-e^{-(\alpha_G+\alpha)}}\right)\right\}\;.
 \nonumber
 \ea
 Using the above expressions, the quantity $M^d$ is found to be:
 \ba
 \label{eq:Md}
 M^d &=& g^2 C_F \int \frac{d^4 l}{(2\pi)^4}
 \sum_\mu \kappa^2_\mu \Biggl[ \cos^2 l_\mu/2
 \left((e^{-\alpha}-b)\widetilde{G^M}+R_0^M+R_1^M\right)\nonumber \\
 &-&\sin^2 l_\mu/2\left((e^{-\alpha}-b)\widetilde{G^M}+R_0^M+R_2^M\right)+
   \kappa_\mu \sin^2 l_\mu \widetilde{G^M}
  \Biggr]\nonumber \\
 &\times&\frac{\chi\hat{l}^2_\mu+
 \frac{f^{\mu\mu}(l)}{f_D(l)}}{(\chi^2\hat{l}^2_0+\sum_j\hat{l}^2_j)^2} \;.
 \ea
 Similarly, in the massless case,
 the expression for $M^d_{\rm finite}$ reads:
 \ba
 \label{eq:Md_finite}
 M^d_{\rm finite}=M^d(m=0) -2g^2 C_F
  \int \frac{d^4 l}{(2\pi)^4}\frac{4\chi\theta
 (\pi^2-(\chi^2l^2_0+\sum_j l^2_j))}
 {(\chi^2l^2_0+\sum_j l^2_j)^2}
 \ea
 Using the definition~(\ref{eq:Idtad&Idlog_def}),
 the subtracted part, i.e. $M^d_{\rm log}$, is given
 by Eq.~(\ref{eq:Mlog}) in appendix~\ref{asec:longformulae}.

 Once the values of $I^d_{\rm tad}$, $M^d_{\rm tad}$
 $I^d_\mu(m)$ and  $M^d(m)$ ($I^d_{\rm finite,\mu}$
 and $M^d_{\rm finite}$ in the massless case) are obtained numerically,
 Eq.~(\ref{eq:tildeZ2}) and  Eq.~(\ref{eq:tildeW2})
 (Eq.~(\ref{eq:tildeZ3}) and  Eq.~(\ref{eq:tildeW3}) in the
 massless case) then give all
 the relevant renormalization factors for the chiral
 fermion mode to one-loop order.

 \section{Numerical results for the one-loop calculation}

 In this section, we present numerical results of our
 calculation of the fermion propagator to one-loop.
 We will present the results for the wave function
 renormalization for the light fermions, the mass
 renormalization for the massive quark and the
 shift in the parameter $M_5$ which is important
 in order to have chiral modes.

 \subsection{Wave function renormalization for the domain wall fermion}

 In this section, we concentrate on the wave function
 renormalization constant associated with the chiral mode of
 the domain wall fermion. We will be discussing the massless case and the
 massive case separately.

 For the massless case,
 the formula for the wave-function renormalization
 constant is (c.f. Eq.~(\ref{eq:tildeZ3})):
 \be
 \label{eq:tildeZ4}
 \tilde{Z}_{\mu} = 1+g^2\left(I^d_{\rm tad,\mu}
 + I^d_{\rm log,\mu} +I^d_{\rm finite, \mu}\right)\;,
 \ee
 where $I^d_{\rm tad,\mu}$ comes from tadpole diagram
 and $I^d_{\rm log,\mu}$ while $I^d_{\rm finite,\mu}$
 designates the logarithmic divergent
 and finite part of the half circle diagram, respectively.
 The quantity $I^d_{\rm log,\mu}$, as defined in
 Eq.(\ref{eq:Idtad&Idlog_def}) and Eq.~(\ref{eq:Ilog}),
 can be evaluated analytically. The value of $I^d_{\rm finite,\mu}$
 is obtained from Eq.~(\ref{eq:Idmu_finite}) by numerical integration.
 Similarly, using Eq.~(\ref{eq:Idmu}) one finds
 the value of $I^d_\mu(m)$ in the massive case
 once the bare parameters in the fermion action are given.
 The value of $I^d_{\rm tad,\mu}$ is
 obtained by performing the integration appearing
 in Eq.~(\ref{eq:IMtad}) numerically.

 In Table~\ref{tab:Idtad}, we have listed
 the tadpole diagram contributions to $I^d_{\rm tad,\mu}$
 for several values of anisotropy.
 \begin{table}[tb]
 \caption{Values of $I^d_{\rm tad,\mu}$ for three
 different values of $\chi$ are listed.
  Due to symmetry, $I^d_{\rm tad,i}=I^d_{\rm tad,s}$
 for $i=1,2,3$ and $I^d_{\rm tad,0}=I^d_{\rm tad,t}$.
 \label{tab:Idtad}}
 \begin{center}
 \begin{tabular}{|c|c|c|c|c|c|}
 \hline
 \multicolumn{2}{|c|}{$\chi=1$} & \multicolumn{2}{|c|}{$\chi=3$} &
 \multicolumn{2}{|c|}{$\chi=5$} \\
 \hline
 $I^d_{\rm tad,t}$ & $I^d_{\rm tad,s}$ & $I^d_{\rm tad,t}$ &
 $I^d_{\rm tad,s}$ & $I^d_{\rm tad,t}$ & $I^d_{\rm tad,s}$ \\
 \hline
 $-0.090487$ & $-0.088871$ & $-0.014293$ & $-0.122173$ & $-0.005399$ & $-0.127386$ \\
 \hline
 \end{tabular}
 \end{center}
 \end{table}
 Due to symmetry, we have denoted $I^d_{{\rm tad},i}=I^d_{\rm tad,s}$
 for $i=1,2,3$ and $I^d_{\rm tad,0}=I^d_{\rm tad,t}$.
 The values in Table~\ref{tab:Idtad} are obtained for
 $\chi=1$, $3$ and $5$. Note that the tadpole contribution
 $I^d_{\rm tad,\mu}$ only depends on the anisotropy
 parameter $\chi$ and is independent of other parameters
 in the fermion action.

 \begin{figure}[htb]
 \begin{center}
 \includegraphics[width=12.0cm,angle=0]{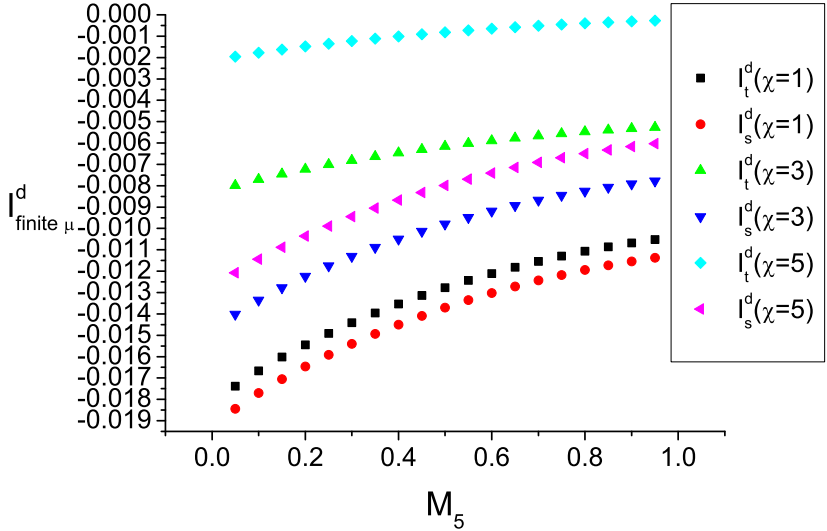}
 \end{center}
 \caption{The values of $I^d_{\rm finite,\mu}$ as a function
 of $M_5$ for three values of anisotropy $\chi=1,3,5$ at
 $\kappa_s=1.0$, $\kappa_t=\chi\kappa_s$.
 \label{fig:IdKs1.0m0.0}}
 \end{figure}
 \begin{figure}[htb]
 \begin{center}
 \includegraphics[width=12.0cm,angle=0]{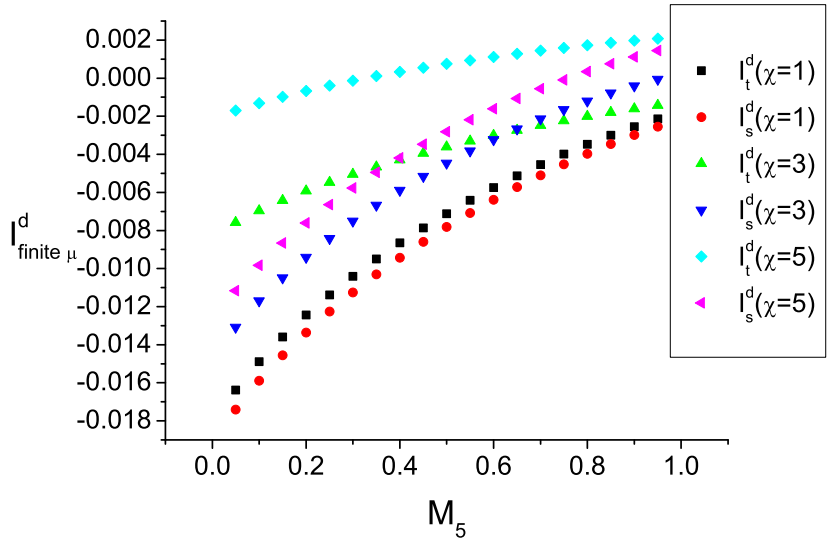}
 \end{center}
 \caption{Same as in Fig.~\ref{fig:IdKs1.0m0.0} but with
 $\kappa_s=0.5$.
 \label{fig:IdKs0.5m0.0}}
 \end{figure}
 For the massless case, the finite part of the half-circle diagram
 contribution, i.e. $I^d_{\rm finite,\mu}$, are calculated using
 numerical integration when the parameters of the theory are
 fixed. In Fig.~\ref{fig:IdKs1.0m0.0} and Fig.~\ref{fig:IdKs0.5m0.0}
 we have plotted the values of $I^d_{\rm finite,\mu}$ as a
 function of the parameter $M_5$ for three values of the
 anisotropy ($\chi=1,3,5$) and two values of $\kappa_s$
 ($\kappa_s=1.0,0.5$). Since $I^d_{\rm finite,\mu}$ always
 corresponds to the massless case, we have set
 $\kappa_t=\chi\kappa_s$ in these plots.
 The infra-red divergent contribution from the half-circle
 diagram can be obtained from Eq.~(\ref{eq:Ilog}) once a
 infra-red scale is given.

 \begin{figure}[htb]
 \begin{center}
 \includegraphics[width=12.0cm,angle=0]{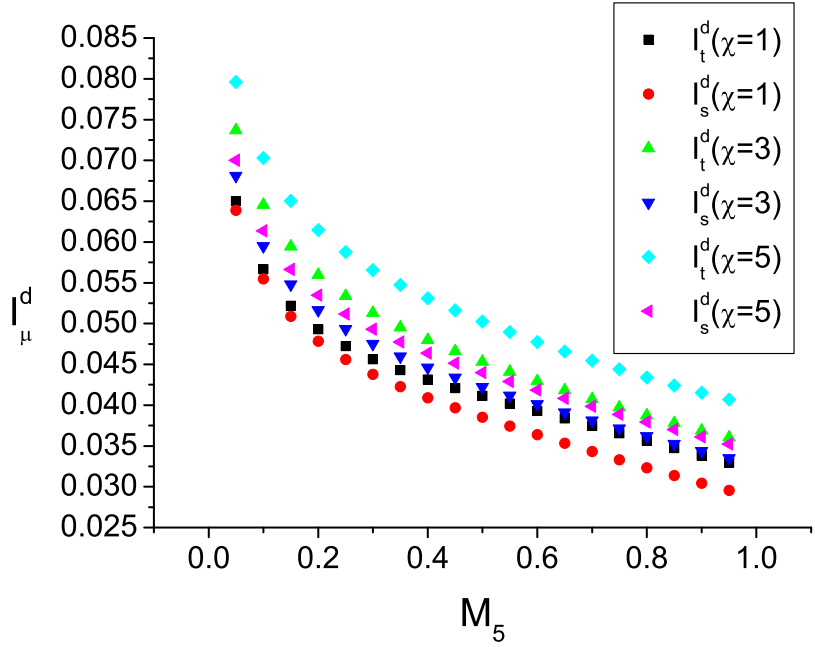}
 \end{center}
 \caption{The values of $I^d_\mu(m)$ as a function of
 $M_5$ for three values of anisotropy: $\chi=1,3,5$
 at current quark mass $m=0.3$. The parameter $\kappa_s=1.0$
 and the values of $\kappa_t$ are adjusted accordingly,
 as discussed in Sec.~\ref{sec:free_G}.
 \label{fig:IdKs1.0m0.3}}
 \end{figure}
 \begin{figure}[htb]
 \begin{center}
 \includegraphics[width=12.0cm,angle=0]{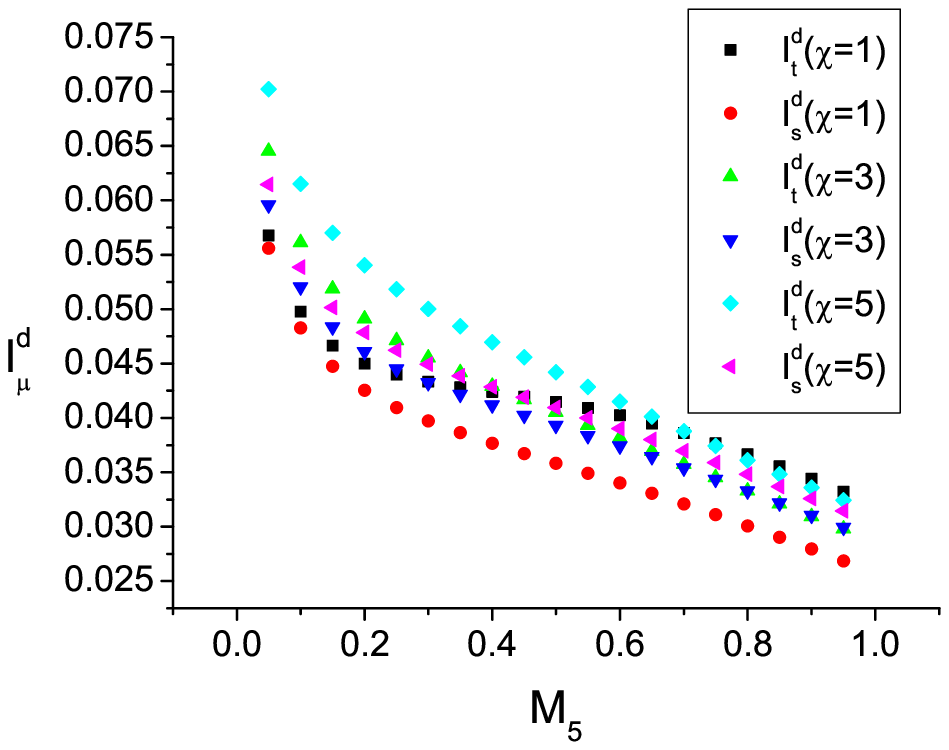}
 \end{center}
 \caption{Same as in Fig.~\ref{fig:IdKs1.0m0.3}
 with $\kappa_s=0.5$ and $m=0.3$.
 \label{fig:IdKs0.5m0.3}}
 \end{figure}
\begin{figure}[htb]
 \begin{center}
 \includegraphics[width=12.0cm,angle=0]{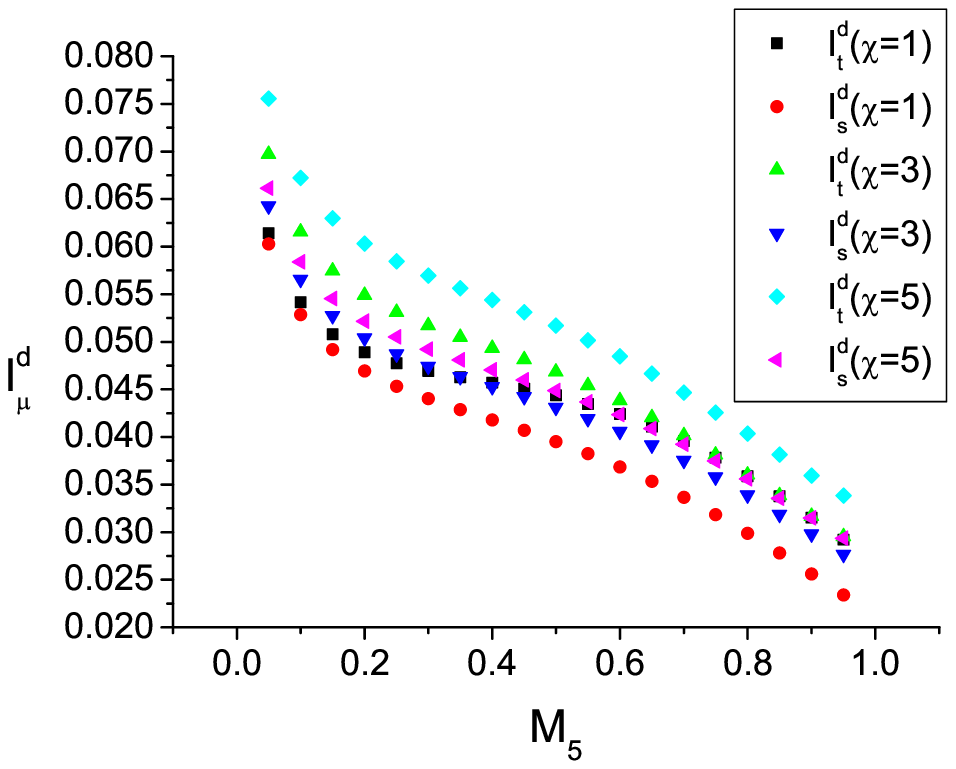}
 \end{center}
 \caption{Same as Fig.~\ref{fig:IdKs1.0m0.3}
 with current quark mass parameter $m=0.5$ and $\kappa_s=1.0$.
 \label{fig:IdKs1.0m0.5}}
 \end{figure}
\begin{figure}[htb]
 \begin{center}
 \includegraphics[width=12.0cm,angle=0]{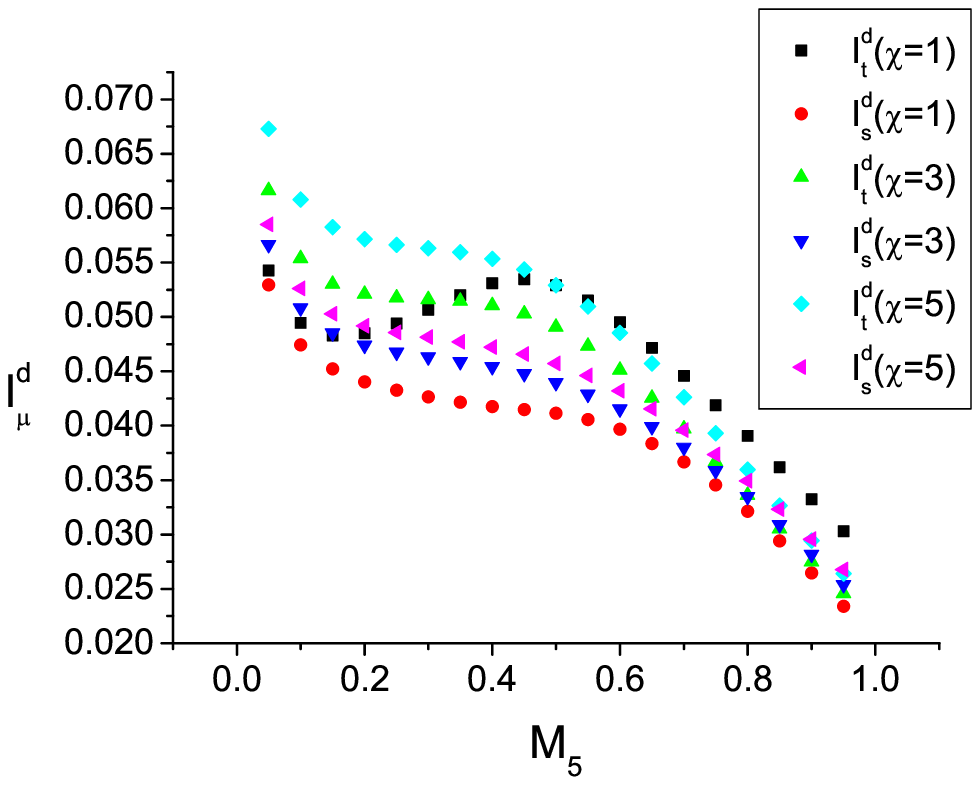}
 \end{center}
 \caption{Same as Fig.~\ref{fig:IdKs1.0m0.3} with
 current quark mass parameter $m=0.5$ and $\kappa_s=0.5$.
 \label{fig:IdKs0.5m0.5}}
 \end{figure}

 For the massive case, one needs the values of $I^d_{\mu}(m)$
 which is obtained by numerically integrating Eq.~(\ref{eq:Idmu}).
 The values of $I^d_\mu(m)$ are shown in Fig.~\ref{fig:IdKs1.0m0.3}
 through Fig.~\ref{fig:IdKs0.5m0.5}.
 Similar to the massless case, the corresponding values of
 $I^d_\mu(m)$ are plotted as functions of $M_5$ for three
 values of anisotropy: $\chi=1,3,5$ and two values of
 $\kappa_s$ ($\kappa_s=1.0,0.5$).

 \subsection{Mass renormalization for the domain wall fermion}

 For the mass renormalization
 factor given by Eq.~(\ref{eq:tildeW3}), we obtain:
 \be
 \label{eq:Zm0}
 \tilde{Z}_m=1+g^2(M^d_{\rm tad}+M^d_{\rm log}+M^d_{\rm finite})\;,
 \ee
 for the massless case and for the massive case we have:
 \be
 \label{eq:Zm1}
 \tilde{Z}_m=1+g^2(M^d_{\rm tad}+M^d)\;.
 \ee
 For the tadpole contribution $M^d_{\rm tad}$, using
 notation~(\ref{eq:Mdtad})
 and setting $\kappa_s=1.0$, we have evaluated
 the tadpole contribution $M^0_{\rm tad}$
 for three values of anisotropy and the results are
 listed in Table~\ref{tab:M0tad}.
 \begin{table}[htb]
 \caption{Values of $M^0_{\rm tad}$ for
 $\chi=1$, $3$, $5$ and $m=0$, $0.3$, $0.5$.
 The values listed here correspond to those
 at $\kappa_s=1$, $M_5=0.5$ and the values of $\kappa_t$ are
 tuned accordingly.
 \label{tab:M0tad}}
 \begin{center}
 \begin{tabular}{|c|c|c|c|}
 \hline
 & $\chi=1$ & $\chi=3$ & $\chi=5$ \\
 \hline
 $m=0.0$   & $-0.357099$ & $-0.409400$ & $-0.409156$ \\
 \hline
 $m=0.3$ & $-0.355505$ & $-0.408316$ & $-0.408381$ \\
 \hline
 $m=0.5$ & $-0.352263$ & $-0.406204$ & $-0.406878$ \\
 \hline
 \end{tabular}
 \end{center}
 \end{table}
 The values of $M^0_{\rm tad}$ listed in
 Table~\ref{tab:M0tad} are obtained at:
 $\kappa_s=1$, $M_5=0.5$ with the values of $\kappa_t$ being
 tuned accordingly.

\begin{figure}[htb]
 \begin{center}
 \includegraphics[width=12.0cm,angle=0]{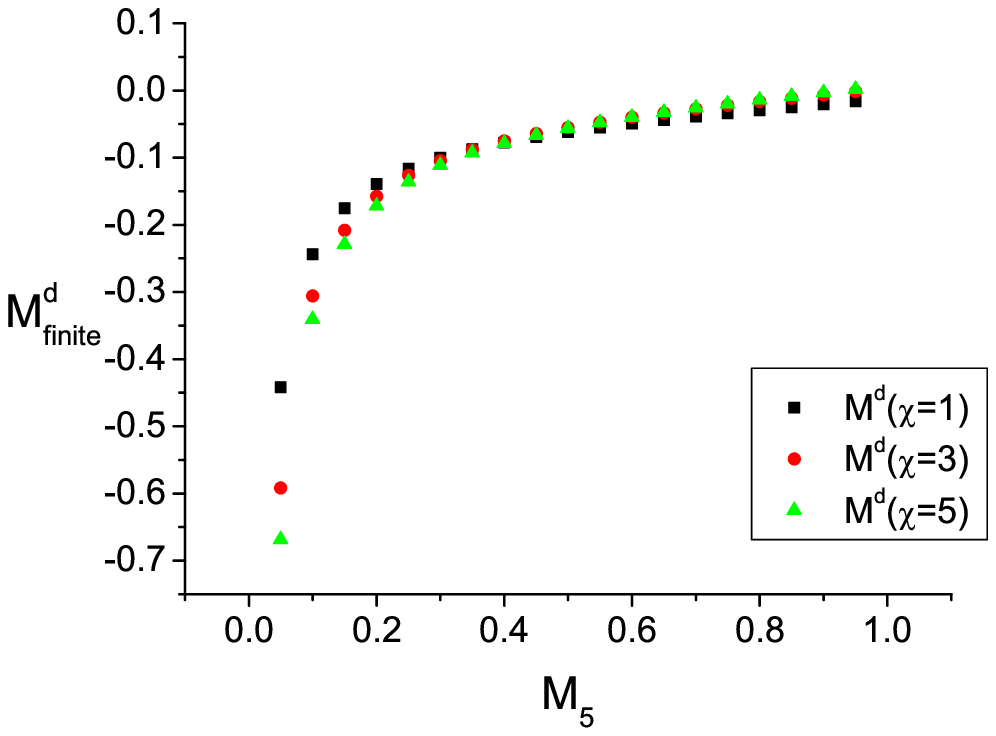}
 \end{center}
 \caption{The values of $M^d_{\rm finite}$ as a function of $M_5$
 are shown for $\chi=1$, $3$ and $5$. Other parameters are:
 $\kappa_s=1$, $\kappa_t=\chi\kappa_s$, $m=0$.
 \label{fig:MdKs1.0m0.0}}
 \end{figure}

\begin{figure}[htb]
 \begin{center}
 \includegraphics[width=12.0cm,angle=0]{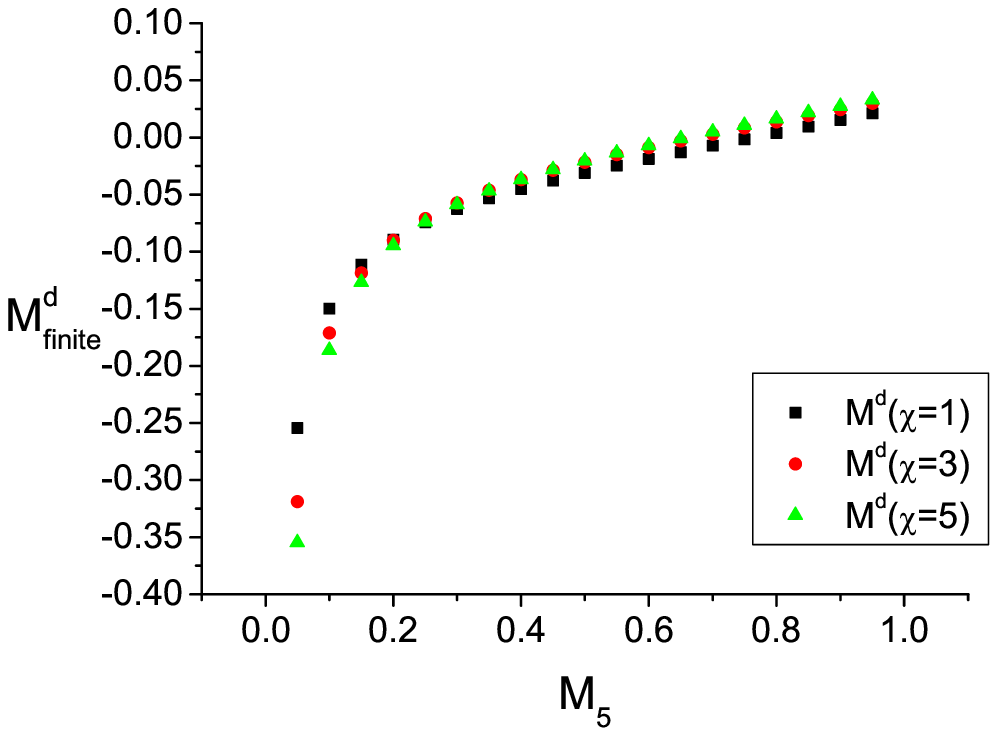}
 \end{center}
 \caption{Same as Fig.~\ref{fig:MdKs1.0m0.0} except
 that $\kappa_s=0.5$.
 \label{fig:MdKs0.5m0.0}}
 \end{figure}

 For the half-circle contributions in the massless case,
 values of $M^d_{\rm finite}$ are presented in
 Fig.~\ref{fig:MdKs1.0m0.0} and Fig.~\ref{fig:MdKs0.5m0.0}
 where we have shown the values of $M^d_{\rm finite}$
 for $\kappa_s=1.0$ and
 $\kappa_s=0.5$, respectively. In each figure, three values
 of the anisotropy, namely  $\chi=1,3,5$, are taken to evaluate
 $M^d_{\rm finite}$ as a function of $M_5$ which is taken to
 vary between zero and unity.

 For the massive case, we evaluate $M^d(m)$ directly
 by numerical integration.
 \begin{figure}[htb]
 \begin{center}
 \includegraphics[width=12.0cm,angle=0]{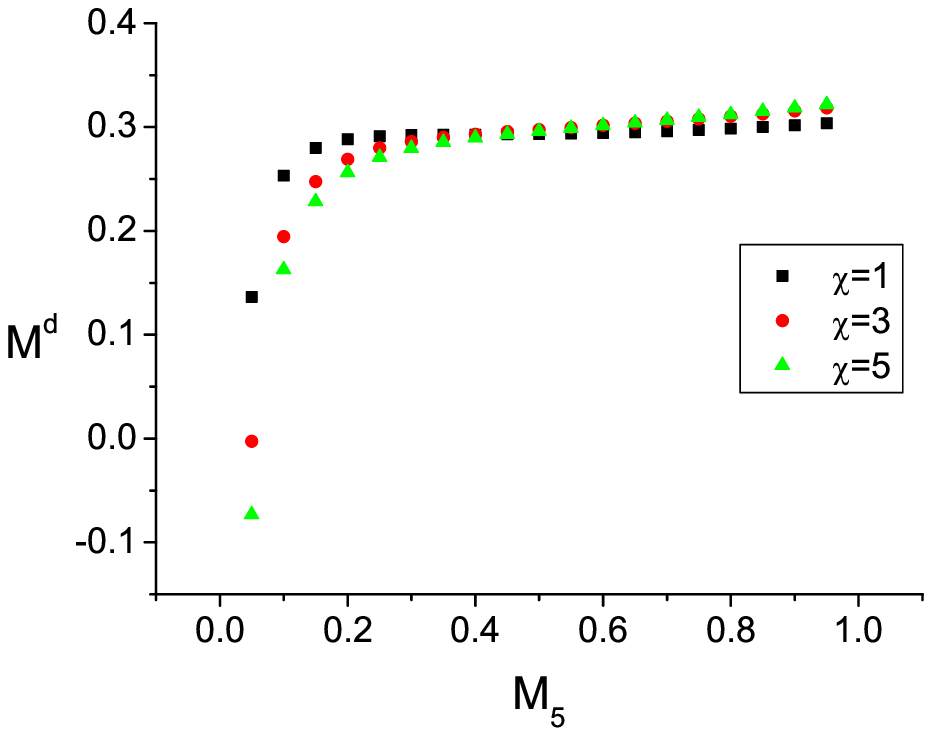}
 \end{center}
 \caption{The value of $M^d$ as a function of $M_5$ is
 shown for $\chi=1$, $3$ and $5$. Other parameters are:
 $\kappa_s=1$,  $m=0.3$.
 \label{fig:MdKs1.0m0.3}}
 \end{figure}
 \begin{figure}[htb]
 \begin{center}
 \includegraphics[width=12.0cm,angle=0]{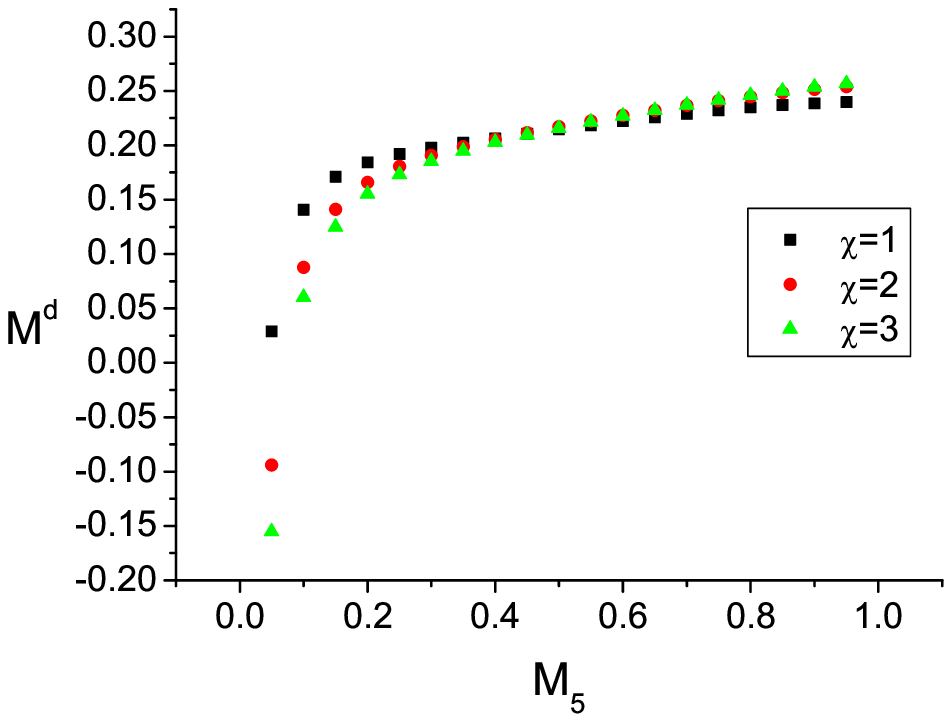}
 \end{center}
 \caption{The value of $M^d$ as a function of $M_5$ is
 shown for $\chi=1$, $3$ and $5$. Other parameters are:
 $\kappa_s=1$,  $m=0.5$.
 \label{fig:MdKs1.0m0.5}}
 \end{figure}
 In Fig.~\ref{fig:MdKs1.0m0.3} and Fig.~\ref{fig:MdKs1.0m0.5},
 we have shown the values of $M^d(m)$ as a function
 of $M_5$ at $m=0.3$ and $m=0.5$, respectively.
 We have also taken three values for the anisotropy
 parameter in each case.

 \subsection{Mass shift for the parameter $M_5$ in mean-field theory}

 At the tree level, it is known that the domain wall
 fermion preserves chiral properties only when the parameter
 $M_5$ is tuned to the right interval ( $0<M_5<1$ for free
 domain wall fermion).
 Since this parameter enters the action via a five-dimensional mass term,
 one expects that it receives substantial additive renormalization.
 At one-loop level, it is
 important to check that, after renormalization effects are taken into account,
 whether this so-called ``chiral window" for $M_5$ remains there or not.
 In the case of isotropic lattices, it has been
 shown that the window remains in perturbation theory
 and one expects chiral properties
 of the fermion as long as  $M_5$ lies in the right interval.
 We wish to check this point for the anisotropic lattices.

 In perturbation theory, as expected,
 the main contribution for the shift of
 the parameter $M_5$ comes from the tadpole diagram.
 To simplify the analysis, we will only discuss the
 shift of $M_5$ due to tadpole diagrams. Within this
 approximation, one finds that the parameter $M_5$ is
 replaced by:
 \be
 \label{eq:deltaM5}
 \tilde{M_5}=M_5-(\kappa_t(1-U_t)+3\kappa_s(1-U_s))
 \equiv M_5-\delta(M_5)\;,
 \ee
 where the tadpole parameters $U_t$ and $U_s$ are
 given by:
 \ba
 \label{eq:U_t&U_s}
 U_t&=&U_0=1-g^2 C_F \frac{1}{2}\int_{-\pi}^\pi \frac{d^4 l}{(2 \pi)^4}
 \frac{1}{(\chi^2\hat{l}^2_0+\sum_j \hat{l}^2_j)^2}
 (\hat{l}^2_0\chi+\frac{f^{00}}{f_D}) \;,
 \nonumber \\
 U_s&=&U_i=1-g^2 C_F \frac{1}{2}\int_{-\pi}^\pi \frac{d^4 l}{(2 \pi)^4}
 \frac{1}{(\chi^2\hat{l}^2_0+\sum_j \hat{l}^2_j)^2}
 (\hat{l}^2_i\chi+\frac{f^{ii}}{f_D}) \;,
 \ea
 here $g^2$ and $\chi$ are the ``boosted" coupling and
 the ``boosted" anisotropy~\cite{drummond02:aniso_xi}
 which are defined as:
 \ba
 \label{eq:g^2&chi}
 g^2=\frac{2N_c}{\beta_0}U^3_sU_t\;,
 \;\;
 \chi=\chi_0\frac{U_s}{U_t}\;.
 \ea
 Eq.~(\ref{eq:deltaM5}) suggests that in perturbation
 the chiral window remains but it is shifted.

 We will only show results at
 zero current quark mass $m=0$ in which case we have:
 $\kappa_t/\kappa_s=\chi$.
 To get a feeling how large the shift in the parameter
 $M_5$ can get, we take a particular value of $\beta_0$.
 At first, the values of $U_t$ and $U_s$ are set as $U_t=U_s=1$.
 By making use of the
 formula (\ref{eq:g^2&chi}), the values of $g^2$ and $\chi$ are
 obtained and can be substituted into  formula (\ref{eq:U_t&U_s}) to
 obtain a new set of values for $U_t$ and $U_s$.
 The above procedure is repeated iteratively until the values of
 $U_t$ and $U_s$ become stable. The ultimate
 values for  $U_t$ and $U_s$ in perturbation theory and
 the corresponding shift in $M_5$ thus obtained are listed
 in Table~\ref{tab:deltaM5} for $\beta_0=3.0$
 \begin{table}[htb]
 \caption{The values of tadpole improvement parameters
 $U_t$, $U_s$ and corresponding values for $\delta M_5$
 as evaluated from Eq.~(\ref{eq:deltaM5}) at
 $\chi_0=1$ to $\chi_0=5$.
 The bare gauge coupling is $\beta_0=3.0$ and
 the parameter $\kappa_s=1$.
 \label{tab:deltaM5}}
 \begin{center}
 \begin{tabular}{|c|c|c|c|c|c|}
 \hline
 & $\chi_0=1$ & $\chi_0=2$ & $\chi_0=3$ & $\chi_0=4$ & $\chi_0=5$ \\
 \hline
 $U_t$ & $0.889580$ & $0.956099$ & $0.977525$ & $0.986559$ & $0.991122$ \\
 \hline
 $U_s$ & $0.891130$ & $0.861399$ & $0.851613$ & $0.847194$ & $0.844862$ \\
 \hline
 $\delta(M_5)$ &
 $0.437222$ & $0.494908$ & $0.503901$ & $0.504587$ & $0.503253$ \\
 \hline
 \end{tabular}
 \end{center}
 \end{table}
 at five different values of $\chi_0$.

 \begin{figure}[htb]
 \begin{center}
 \includegraphics[width=12.0cm,angle=0]{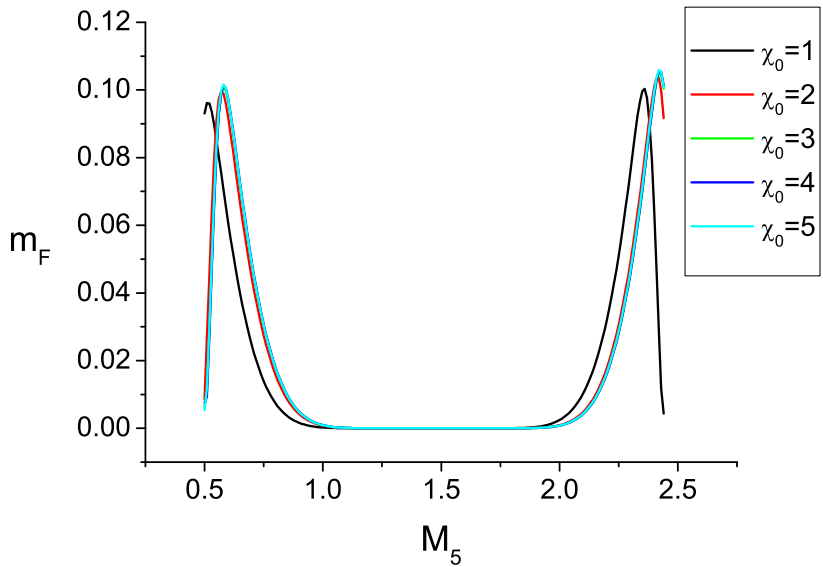}
 \end{center}
 \caption{The effective mass of the chiral mode, $m_F$,
 as a function of the parameter $M_5$ at $m=0.0$, $L_s=10$.
 Five curves with different colors correspond to
 different values of $\chi_0$. The values of $U_t$ and
 $U_s$ are taken to be those in Table~\ref{tab:deltaM5}
 at $\beta_0=3.0$.
 \label{fig:mFvschi0}}
 \end{figure}
 Another way to investigate
 the renormalization of parameter $M_5$ is to use
 mean-field approximation.
 In this case, the domain wall fermion action is rewritten as:
 \ba
 [S_{\rm MF}]_{s,s'}&=&\left(-\tilde{b}(p)+i\sum_\mu \kappa_\mu U_\mu \gamma_\mu
 \sin p_\mu\right)\delta_{s,s'} \nonumber \\
 &+&P_R\delta_{s+1,s'}+P_L\delta_{s-1,s'}\;,
 \ea
 where $U_\mu$ are pure numbers
 \footnote{We use the notation: $U_t=U_0$ and $U_s=U_i$ for
 $i=1,2,3$.}
 and $\tilde{b}(p)$ is defined as:
 \ba
 \tilde{b}(p) &=& 1-M_5+\sum_\mu\kappa_\mu(1-U_\mu \cos p_\mu)
 \;.
 \ea
 Following similar calculations as for the free domain wall
 propagator, one can define an effective mass for the fermion
 in the mean-field approximation. We will denote this
 mass parameter as $m_F$.

 In Fig.~\ref{fig:mFvschi0}, we show the effective mass
 $m_F$ of the chiral mode as a function
 of the parameter $M_5$ for five values of $\chi_0$.
 For simplicity, we have set $\kappa_s=1$ in this figure
 and the extent in the fifth dimension is taken to be $L_s=10$.
 It is seen from the figure that, within a certain interval,
 the chiral window remains for all $\chi_0$.
 This has been checked in the case of isotropic lattice
 which roughly corresponds to the red
 curve in our figure. We see from the figure that
 this qualitative feature remains valid in the anisotropic
 lattice case for almost all values of anisotropy $\chi_0$.
 In fact the shape of the curves is quite insensitive
 to the value of $\chi_0$, as is seen from the figure.
 The size of the chiral window remains more or less
 unchanged but the starting and ending position is
 shifted. The results discussed above suggest that,
 as long as the parameter $M_5$
 is tuned to the right interval,  massless chiral
 modes are stable against perturbative quantum fluctuations.
 This result also serves as a guidance for the choice
 of the parameters in our future numerical simulations.

 \begin{figure}[htb]
 \begin{center}
 \includegraphics[width=12.0cm,angle=0]{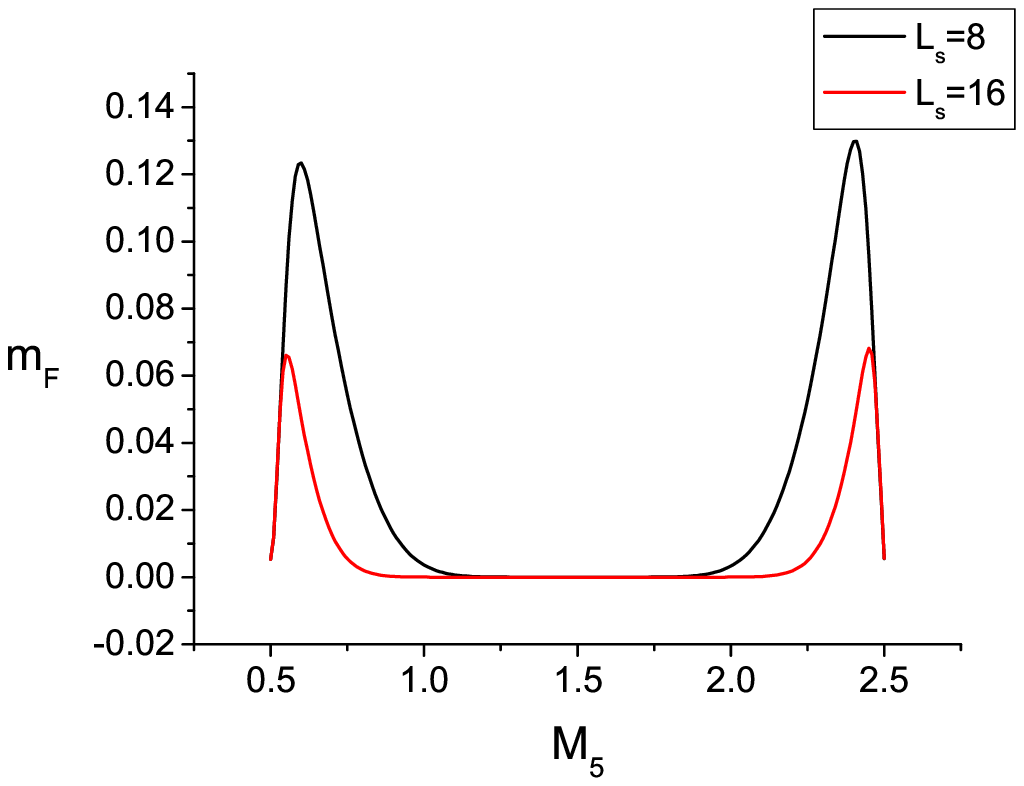}
 \end{center}
 \caption{The effective mass $m_F$ of the chiral mode as a function
 of $M_5$ for two different values of $L_s$. The red
 and black curve corresponds to $L_s=16$ and $L_s=8$,
 respectively. Other parameters are the same as
 those in Fig.~\ref{fig:mFvschi0}.}
 \label{fig:mFvsL}
 \end{figure}
 We have also investigated the dependence of the
 chiral window on the size of the fifth dimension $L_s$.
 In Fig.~\ref{fig:mFvsL}, we have plotted
 the effective mass of the chiral mode as a function of $M_5$ at two
 different values of $L_s$. The anisotropy parameter is set
 to $\chi_0=5$ while all other parameters are kept the
 same as those in Fig.~\ref{fig:mFvschi0}.
 It is seen from the figure that the chiral window for
 the smaller lattice ($L_s=8$) becomes narrower than
 that of the larger lattice ($L_s=16$), as expected.

 \section{Conclusions}
 \label{sec:conclude}

 In this paper we have studied the domain wall fermions on
 anisotropic four-dimensional lattices.
 We have analyzed the free domain wall fermions,
 showing that the hopping parameters have to be tuned
 according to the current quark mass parameter in order to restore the
 correct dispersion relation for the chiral mode of
 the domain wall fermion. Using lattice perturbation
 theory, the domain wall fermion self-energy is calculated to
 one-loop order. We obtained the wave-function renormalization
 constant $\tilde{Z}_\mu$ and the mass renormalization constant
 $\tilde{Z}_m$ for the chiral mode. It is also verified that
 the chiral property of the fermion remains when quantum
 corrections are considered. We have also estimated the range
 for the parameter $M_5$ in order to maintain a chiral mode.

 From our study, it seems that domain wall fermion on
 anisotropic lattices is applicable in practical Monte
 Carlo simulations. The only parameter that has to be tuned
 is the hopping parameter $\kappa_t$ when other parameters
 are given. The tuning of this parameter can be done
 either perturbatively or non-perturbatively.
 The non-perturbative tuning requires the
 measurement of physical hadron dispersion relations
 or quark dispersion relations by Monte Carlo simulations.
 An exploratory numerical study is now under way and
 we hope to come up with some results in the future~\cite{chuan06:future}.
 Compared with similar situations for the Wilson fermions,
 anisotropic lattice domain wall fermion
 appears to be less complicated since there is no need to add
 clover terms (or other dimension five operators)
 due to better chiral properties. Of course, this conclusion
 relies on the fact that the residual mass effects due to
 the finite extent of the fifth dimension to be small.
 We have tried to estimate this effect with mean-field
 approximation but this issue definitely should  be checked further
 in future numerical studies.

 It should be straightforward to generalize the results we
 obtained in this paper to the overlap fermions on anisotropic
 lattices. Another topic that is worth studying is the renormalization
 factors for quark bilinear operators. The study of these
 problems are now under investigation.
 To summarize, we anticipate the anisotropic domain wall fermions
 to be helpful in the study of
 the heavy-light hadronic systems, exotic hadrons (hybrid) with light quarks
 and light hadron-hadron scattering processes and we hope to
 come up with some exploratory numerical results in the future~\cite{chuan06:future}.

 \appendix
 \section{Dispersion relation of free domain wall fermions}
  \label{asubsec:aniso}

 In this appendix, we briefly outline the derivation of
 the dispersion relation for the free domain wall fermion on
 anisotropic lattices.
 The dispersion relation is such that the four-momentum
 $p=(iE_\bp,\bp)$ is a zero of the function:
 \be
 \Delta=(1+mm_r)^2be^{2\alpha_G}
 -(1-m^2)(1-m_r^2)e^{\alpha_G}-(m+m_r)^2b\;.
 \ee
 In principle, the parameter $m_r=e^{-\alpha_GL_s}$ also
 depends on the four-momentum. For the moment, we assume
 that the extension in the fifth dimension is large enough
 such that we can ignore the momentum dependence of
 $m_r$ in the study of the dispersion
 relation. This amounts to taking $m_r=e^{-\alpha_G(0)L_s}$.
 If we denotes:
 \be
 m_1=(1+mm_r)^2\;,\;\;
 m_2=(1-m^2)(1-m^2_r)\;,
 m_3=(m+m_r)^2\;.
 \ee
 Then, the equation for the pole yields:
 \be
 \label{eq:simplified}
 -\tilde{p}^2=M_1\left[(1+b^2+\tilde{p}^2)^2-4b^2\right]
 \;,
 \ee
 where $M_1=m_1m_3/m_2^2$. The complete
 dispersion relation turns out to be quite complicated. But if
 the lattice three-momentum $|\bp|\ll 1$, we obtain:
 \be
 E_\bp=E_0+{\bp^2\over 2M_{kin}}+ O(\bp^4)\;,
 \ee
 where $E_0$ will be identified as the pole
 mass of the quark: $E_0=m_Q$ and $M_{kin}$ is
 the so-called kinetic mass.
 After some calculations, the equation satisfied
 by $E_0=m_Q$ is found to be:
 \footnote{Note that the quantity $E_0$ appearing
 in this equation is the dimensionless quantity:
 $a_tE_0=a_tm_Q=(a_sm_Q)/\chi$,
 if we restore the dimension for $E_0$.}
 \be
 \label{eq:mQ}
 A\cosh^2E_0+B\cosh E_0+C=0
 \ee
 where
 \ba
 A&=&\kappa_t^2-4\kappa_t^2\left[(b(0)+\kappa_t)^2-1\right]M_1
 \nonumber \\
 B&=&4(b(0)+\kappa_t)\kappa_t\left[(b(0)+\kappa_t)^2+\kappa_t^2-1\right]M_1
 \nonumber \\
 C&=&-\kappa_t^2-\left[(b(0)+\kappa_t)^4+2(b(0)+\kappa_t)^2(\kappa_t^2-1)+
 (\kappa_t^2+1)^2\right]M_1
 \ea
 We can also find out the kinetic mass term with
 the result:
 \be
 {1\over 2M_{kin}}=\frac{d E}{d\vec{p}^2}\left|_{\vec{p}=0}=\frac{A_1\cosh^2E_0+B_1\cosh E_0+C_1}
 {(2A\cosh E_0+B)\sinh E_0}\right.
 \ee
 where
 \ba
 A_1&=&4(b(0)+\kappa_t)\kappa_t^2\kappa_sM_1
 \nonumber \\
 B_1&=&-2\left[3(b(0)+\kappa_t)^2+2(b(0)+\kappa_t)\kappa_s+\kappa_t^2-1\right]
 \kappa_t\kappa_sM_1
 \nonumber \\
 C_1&=&\kappa_s^2+2\kappa_s\left[(b(0)+\kappa_t)\left((b(0)+\kappa_t)^2+\kappa_t^2-1\right)
 \right.\nonumber \\
 &+&\left.\kappa_s\left((b(0)+\kappa_t)^2+\kappa_t^2+1\right)\right]M_1
 \ea
 In order to have the usual energy-momentum
 dispersion relation, one imposes the condition:
 $M_{kin}=m_Q$ which yields the relationship between
 $\chi$ and $\kappa_t,\kappa_s$ as
 (restoring the lattice spacing explicitly):
 \be
 \label{eq:chi}
 \chi^2=\frac{(2A\cosh E_0+B)\sinh E_0}{2(A_1\cosh^2E_0+B_1\cosh E_0+C_1)E_0}
 \;,
 \ee
 where $E_0=m_Q$ is obtained by solving Eq.~(\ref{eq:mQ}).
 Eq.~(\ref{eq:mQ}) and Eq.~(\ref{eq:chi}) thus establish the relation
 among the pole mass $m_Q$, anisotropy $\chi$ and the other
 bare parameters of the theory for free domain wall fermions
 on anisotropic lattices.

 In the above discussions, we have neglected the momentum
 dependence of $m_r=e^{-\alpha_G(p)L_s}$. If we keep this
 momentum dependence, the pole position can be solved numerically.
 For any given set of parameters, we start with the choice
 $m_r=e^{-\alpha_G(0)L_s}$ and solve for $m_Q$ in
 Eq.~(\ref{eq:mQ}). Then the obtained value of $E_0=m_Q$ is
 substituted into $m_r=e^{-\alpha_G(iE_0,0,0,0)L_s}$ and a
 new value of $E_0$ is thus obtained by solving
 Eq.~(\ref{eq:mQ}) again. This procedure can be iterated
 until a stable set of values for
 $m_Q$ and $m_r=e^{-\alpha_G(im_Q,0,0,0)L_s}$ is obtained.
 The figures in the main text are obtained using this method
 for a given $L_s=8$.

 \section{Some Explicit formulae for loop integrals}
 \label{asec:longformulae}

 In this appendix, we list the explicit expressions
 for various loop integrals which enters the fermion
 self-energy discussed in section~\ref{subsec:self_energy}.

 First, the tadpole contributions $I_{\rm tad,\mu}$ and $M_{\rm tad}$
 are given by:
 \ba
 \label{eq:IMtad}
 I_{\rm tad,\mu} (s, s') &=& -g^2 C_F \frac{1}{2}\int_{-\pi}^\pi \frac{d^4 l}{(2 \pi)^4}
 \frac{1}{(\chi^2\hat{l}^2_0+\sum_j \hat{l}^2_j)^2}
 (\hat{l}^2_\mu\chi+\frac{f^{\mu\mu}}{f_D}) \delta_{s,s'}\;,
 \nonumber
 \\
 M_{\rm tad}(s, s') &=& -g^2 C_F \frac{1}{2}\int_{-\pi}^\pi \frac{d^4 l}{(2 \pi)^4}
 \sum_\mu\frac{\kappa_\mu}{(\chi^2\hat{l}^2_0+\sum_j \hat{l}^2_j)^2}
 (\hat{l}^2_\mu\chi+\frac{f^{\mu\mu}}{f_D}) \delta_{s,s'}\;.
 \ea
 Note that these tadpole contributions are diagonal in
 the fifth dimension. It is also noted that
 $I_{\rm tad,\mu}$ and $M_{\rm tad}$ are independent
 of the current quark mass parameter $m$ explicitly.

 Now comes the formulae for the half-circle diagrams.
 In massive case, we have:
 \ba
 \label{eq:Ifinite_plus_massive}
 [I_\mu^+(m)]_{s,s'} &=& g^2 C_F \int \frac{d^4 l}{(2\pi)^4}
 \Biggl\{\Biggl[ \frac{1}{2}\kappa_\mu\left( \cos l_\mu (
 W^-G_R + W^+G_L)(s,s')\right. \nonumber
 \\
 &+&\left. \kappa_\mu\sin^2l_\mu (G_L+G_R)(s,s')
 \right)\Biggr]\frac{\chi\hat{l}^2_\mu+
 \frac{f^{\mu\mu}(l)}{f_D(l)}}{(\chi^2\hat{l}^2_0+\sum_j\hat{l}^2_j)^2}\nonumber
 \\
 &+&\Biggl[\left(\frac{1}{2}\kappa_\mu
 ( W^-G_R + W^+G_L)-2\kappa^2_\mu\cos^2 l_\mu/2
 G_L\right)(s,s')\times \nonumber \\
 & &\left(\frac{4(\chi^2)^{\delta_{\mu0}}}{\chi^2\hat{l}^2_0+\sum_j\hat{l}^2_j}
 (\chi\hat{l}^2_\mu+\frac{f^{\mu\mu}(l)}{f_D(l)})-2\chi
 -\frac{f^\mu_\mu(l)}{f_D(l)}+\frac{f^D_\mu(l)f^{\mu\mu}(l)}{f^2_D(l)}\right)\nonumber
 \\
 &-&2\chi\kappa^2_\mu
 \left(\sin^2 l_\mu/2 G_R+\cos^2 l_\mu/2 G_L\right)(s,s')\nonumber
 \\
 &+&\sum_\nu \kappa^2_\nu
 \left(\sin^2 l_\nu/2 G_R+\cos^2 l_\nu/2 G_L\right)(s,s') \nonumber
 \\
 &\times&\left(\frac{4(\chi^2)^{\delta_{\mu0}}}{\chi^2\hat{l}^2_0+\sum_j\hat{l}^2_j}
 (\chi\hat{l}^2_\nu+\frac{f^{\nu\nu}(l)}{f_D(l)})-\frac{f^\nu_\mu(l)}{f_D(l)}+
 \frac{f^D_\mu(l)f^{\nu\nu}(l)}{f^2_D(l)}\right)\Biggr]\nonumber
 \\
 &\times&\frac{\sin^2
 l_\mu}{(\chi^2\hat{l}^2_0+\sum_j\hat{l}^2_j)^2}\Biggr\}
 \\
 \label{eq:Ifinite_minus_massive}
 [I_\mu^-(m)]_{s,s'} &=& g^2 C_F \int \frac{d^4 l}{(2\pi)^4}
 \Biggl\{\Biggl[ \frac{1}{2}\kappa_\mu\left( \cos l_\mu (
 W^-G_R + W^+G_L)(s,s')\right. \nonumber
 \\
 &+&\left. \kappa_\mu\sin^2l_\mu (G_L+G_R)(s,s')
 \right)\Biggr]\frac{\chi\hat{l}^2_\mu+
 \frac{f^{\mu\mu}(l)}{f_D(l)}}{(\chi^2\hat{l}^2_0+\sum_j\hat{l}^2_j)^2}
 \nonumber \\
 &+&\Biggl[\left(\frac{1}{2}\kappa_\mu
 ( W^-G_R + W^+G_L)-2\kappa^2_\mu\cos^2 l_\mu/2
 G_R\right)(s,s')\nonumber \\
 &\times&\left(\frac{4(\chi^2)^{\delta_{\mu0}}}{\chi^2\hat{l}^2_0+\sum_j\hat{l}^2_j}
 (\chi\hat{l}^2_\mu+\frac{f^{\mu\mu}(l)}{f_D(l)})-2\chi
 -\frac{f^\mu_\mu(l)}{f_D(l)}+\frac{f^D_\mu(l)f^{\mu\mu}(l)}{f^2_D(l)}\right)\nonumber
 \\
 &-&2\chi\kappa^2_\mu
 \left(\sin^2 l_\mu/2 G_L+\cos^2 l_\mu/2 G_R\right)(s,s')\nonumber
 \\
 &+&\sum_\nu \kappa^2_\nu
 \left(\sin^2 l_\nu/2 G_L+\cos^2 l_\nu/2 G_R\right)(s,s') \nonumber
 \\
 &\times&\left(\frac{4(\chi^2)^{\delta_{\mu0}}}{\chi^2\hat{l}^2_0+\sum_j\hat{l}^2_j}
 (\chi\hat{l}^2_\nu+\frac{f^{\nu\nu}(l)}{f_D(l)})-\frac{f^\nu_\mu(l)}{f_D(l)}+
 \frac{f^D_\mu(l)f^{\nu\nu}(l)}{f^2_D(l)}\right)\Biggr]\nonumber
 \\
 &\times&\frac{\sin^2
 l_\mu}{(\chi^2\hat{l}^2_0+\sum_j\hat{l}^2_j)^2}\Biggr\}
 \\
 \label{eq:Mfinite_plus_massive}
 [M^+(m)]_{s,s'} &=& g^2 C_F \int \frac{d^4 l}{(2\pi)^4}
 \sum_\mu \kappa^2_\mu \Biggl[ \cos^2 l_\mu/2
 (W^+G_L)(s,s')\nonumber \\
 &-&\sin^2 l_\mu/2(W^-G_R)(s,s')+
  \frac{1}{2} \kappa_\mu \sin^2 l_\mu ( G_L+G_R)(s,s')
  \Biggr]\nonumber \\
 &\times&\frac{\chi\hat{l}^2_\mu+
 \frac{f^{\mu\mu}(l)}{f_D(l)}}{(\chi^2\hat{l}^2_0+\sum_j\hat{l}^2_j)^2}
 \\
 \label{eq:Mfinite_minus_massive}
 [M^-(m)]_{s,s'} &=& g^2 C_F \int \frac{d^4 l}{(2\pi)^4}
 \sum_\mu \kappa^2_\mu \Biggl[ \cos^2 l_\mu/2
 (W^-G_R)(s,s')\nonumber \\
 &-&\sin^2 l_\mu/2(W^+G_L)(s,s')+
  \frac{1}{2} \kappa_\mu \sin^2 l_\mu ( G_L+G_R)(s,s')
  \Biggr]\nonumber \\
 &\times&\frac{\chi\hat{l}^2_\mu+
 \frac{f^{\mu\mu}(l)}{f_D(l)}}{(\chi^2\hat{l}^2_0+\sum_j\hat{l}^2_j)^2}
 \;,
 \ea
 where we have used the following notations:
 \be
 \label{eq:fmunu}
 f^\nu_\mu(l) = -\frac{\partial f^{\mu\mu}(p-l)}{\sin l_\mu \partial
 p_\mu}\Bigg|_{p=0}, \quad
 f^D_\mu(l) =-\frac{\partial f_D (p-l)}{\sin l_\mu \partial
 p_\mu}\Bigg|_{p=0}\;.
 \ee
 It's easy to verify that $M^\pm(m)$ is proportional to
 $m_P=(1-b^2(0))m$. Thus, we may define:
 \ba
 [M^\pm(m)]_{s,s'}=(1-b^2(0))m[M_0^\pm(m)]_{s,s'}
 \ea
 In massless case:
 \ba
 \label{eq:Ilog}
 I^\pm_{\log,\mu} (s,s') &=& \frac{1}{16\pi^2} g^2 C_F
 (C_\pm)_{s,s'}\Biggl( \ln (\pi^2) + 1 -\ln(\tilde{p}^2)+\ln{\kappa^2_s} \Biggr)
 \\
 \label{eq:Ifinite_plus_massless}
 I_{\rm finite,\mu}^+ (s,s') &=&[I_\mu^+(m=0)]_{s,s'}-g^2 C_F
 (C_+)_{s,s'} \nonumber \\
 &\times&\int \frac{d^4 l}{(2\pi)^4}
 \frac{4\chi(\chi^2)^{\delta_{\mu0}}l^2_\mu\theta
 (\pi^2-(\chi^2l^2_0+\sum_j l^2_j))}
 {(\chi^2l^2_0+\sum_j l^2_j)^3}
 \\
 \label{eq:Ifinite_minus_massless}
 I_{\rm finite,\mu}^- (s,s') &=&[I_\mu^-(m=0)]_{s,s'}-g^2 C_F
 (C_-)_{s,s'} \nonumber \\
 &\times&\int \frac{d^4 l}{(2\pi)^4}
 \frac{4\chi(\chi^2)^{\delta_{\mu0}}l^2_\mu\theta
 (\pi^2-(\chi^2l^2_0+\sum_j l^2_j))}
 {(\chi^2l^2_0+\sum_j l^2_j)^3}
 \\
 \label{eq:Mlog}
 M^\pm_{\log} (s,s')&=&\frac{1}{4\pi^2} g^2 C_F
 (C^M_\pm)_{s,s'} \left( \ln (\pi^2) +1 -\ln(\tilde{p}^2)+\ln{\kappa^2_s} \right)
 \\
 \label{eq:Mfinite_plus_massless}
 M^+_{\rm finite} (s,s') &=&m_P
 \left\{[M_0^+(m=0)]_{s,s'}\right.\nonumber \\
 &-&g^2 C_F\left. \frac{(C^M_+)_{s,s'}}{m_P} \int \frac{d^4 l}{(2\pi)^4}
 \frac{4\chi\theta
 (\pi^2-(\chi^2l^2_0+\sum_j l^2_j))}
 {(\chi^2l^2_0+\sum_j l^2_j)^2} \right\}
 \\
 \label{eq:Mfinite_minus}
 M^-_{\rm finite} (s,s') &=&m_P
 \left\{[M_0^-(m=0)]_{s,s'}\right.\nonumber \\
 &-&\left. g^2 C_F \frac{(C^M_-)_{s,s'}}{m_P} \int \frac{d^4 l}{(2\pi)^4}
 \frac{4\chi\theta
 (\pi^2-(\chi^2l^2_0+\sum_j l^2_j))}
 {(\chi^2l^2_0+\sum_j l^2_j)^2} \right\}
 \ea
 where $(C^M_\pm)_{s,s'}$ are given by:
 \ba
 (C^M_+)_{s,s'}&=&(1-b^2(0))m\left[\delta_{s,L_s-1}e^{-\alpha_0s'}\right.
 \nonumber \\
 &+& \left.b(0)\left(\sum_tW^+(0)_{s,t}e^{-\alpha_0(L_s-1-t+s')}\right)\right]
 \;,
 \\
 (C^M_-)_{s,s'}&=&(1-b^2(0))m\left[\delta_{s,0}e^{-\alpha_0(L_s-1-s')}\right.
 \nonumber \\
 &+& \left.b(0)\left(\sum_tW^-(0)_{s,t}e^{-\alpha_0(L_s-1+t-s')}\right)\right]
 \;.
 \ea

 \bibliography{aniso_DWF_perturbative_v2}
 \end{document}